\documentclass[12pt,a4]{paper}
\usepackage{axodraw,a4}
\usepackage{minitoc,cite,fancyhdr}
\usepackage{bbm} 
\usepackage{euscript,amssymb,latexsym} 
\usepackage{epsf,epsfig,graphics,subfigure,graphicx}
\usepackage{amsmath}
\usepackage{multirow}
\usepackage{enumerate}
\usepackage{slashed}
\usepackage{hyperref}

\usepackage{eepic}
             
\makeatletter
\def\fmslash{\@ifnextchar[{\fmsl@sh}{\fmsl@sh[0mu]}}
\def\fmsl@sh[#1]#2{%
  \mathchoice"\quad" latex
    {\@fmsl@sh\displaystyle{#1}{#2}}%
    {\@fmsl@sh\textstyle{#1}{#2}}%
    {\@fmsl@sh\scriptstyle{#1}{#2}}%
    {\@fmsl@sh\scriptscriptstyle{#1}{#2}}}
\def\@fmsl@sh#1#2#3{\m@th\ooalign{$\hfil#1\mkern#2/\hfil$\crcr$#1#3$}}
\makeatother
\global\arraycolsep=2pt 
\def\be{\begin{equation}}
\def\ee{\end{equation}}
\newcommand{\bea}{\begin{eqnarray}}
\newcommand{\eea}{\end{eqnarray}}

\newcommand{\GluonMass}{{\tt{GluonMass}}}
\newcommand{\Qmin}{{\tt{Qmin}}}
\newcommand{\ClSmrLight}{{\tt{ClSmrLight}}}
\newcommand{\ClPowLight}{{\tt{ClPowLight}}}
\newcommand{\pTmin}{{\tt{pTmin}}}
\newcommand{\AlphaMZ}{{\tt{AlphaMZ}}}
\newcommand{\ClMaxLight}{{\tt{ClMaxLight}}}
\newcommand{\PSplitLight}{{\tt{PSplitLight}}}
\newcommand{\PwtDIquark}{{\tt{PwtDIquark}}}
\newcommand{\PwtSquark}{{\tt{PwtSquark}}}

\begin{document}
\begin{flushright}
IPPP/12/46\\
DCPT/12/92\\
MCnet-12-08
\end{flushright}
{\Large \bf Investigation of Monte Carlo Uncertainties on Higgs Boson searches using Jet Substructure}\\[0.7cm]
\hspace{-16pt}{\bf\normalsize{Peter Richardson, David Winn}}
\\ {\it\normalsize{Institute of Particle Physics Phenomenology,
    Department of Physics}} 
\\ {\it\normalsize{University of Durham,
    DH1 3LE, UK;}} 
\\ {\it\normalsize{Email:} }
   {\sf\normalsize{peter.richardson@durham.ac.uk,
       d.e.winn@durham.ac.uk}}
\begin{abstract}

We present an investigation of the dependence of searches for boosted Higgs
bosons using jet substructure on the perturbative and non-perturbative 
parameters of the \textsf{Herwig++} Monte Carlo event generator. Values
are presented for a new tune of the parameters of the event generator,
together with the an estimate of the uncertainties based on varying the
parameters around the best-fit values.

\end{abstract}
\thispagestyle{empty} \thispagestyle{plain} \setcounter{page}{1}
\noindent
---------------------------------------------------------------------------------------------------------
\section{Introduction}

Monte Carlo simulations are an essential tool in the analysis of modern
collider experiments. These event generators contain a large number of both perturbative
and non-perturbative parameters which are tuned to a wide range of
experimental data. While significant effort has been devoted to 
the tuning of the parameters to produce a best fit there has been much less 
effort understanding the uncertainties in these results.
Historically a best fit result, or at best a small
number of tunes, are produced and used to predict observables making
it difficult to assess the uncertainty on any prediction.
The ``Perugia'' tunes\cite{Skands:2010ak,Cooper:2011gk} have 
addressed this by producing a range of tunes by varying specific parameters in the
\textsf{{\sc{Pythia}}}~\cite{Sjostrand:2006za} event generator to produce an  uncertainty. 

Here 
we make use of the \textsf{Professor} Monte Carlo tuning system~\cite{Buckley:2009bj}
to give an assessment of the uncertainty by varying all  the parameters simultaneously 
about the best-fit values by diagonalizing the error matrix. 
This then allows us to systematically estimate the uncertainty on any Monte Carlo
prediction from the tuning of the event generator. We will illustrate
this by considering the uncertainty on jet substructure searches for the Higgs boson
at the LHC.

As the LHC takes increasing amounts of data the discovery of the Higgs
boson is likely in the near future. Once we have discovered the Higgs
boson, most likely in the diphoton channel, it will be vital to
explore other channels and determine if the properties of the observed
Higgs boson are consistent with the Standard Model.  For many years it
was believed that it would be difficult, if not impossible, to observe
the dominant $h^0\rightarrow b\,\bar{b}$ decay mode of a light Higgs
boson.  However, in recent years the use of jet
substructure~\cite{Butterworth:2008iy,Ellis:2009su,Ellis:2009me,Kribs:2009yh,
  Kribs:2010hp,Kribs:2010ii,Butterworth:2009qa,Aaltonen:2011pg,Bai:2011mr,
  Vermilion:2011nm,Plehn:2009rk,Seymour:1993mx,Abdesselam:2010pt,Feige:2012vc,
  Altheimer:2012mn,Yang:2011jk}
offers the possibility of observing this mode. Jet substructure for
$h^0\rightarrow b\,\bar{b}$ as a Higgs boson search channel, was first
studied in Ref.~\cite{Butterworth:2008iy} building on previous work of
a heavy Higgs boson decaying to $W^\pm$ bosons~\cite{Seymour:1993mx},
high-energy $WW$ scattering \cite{Butterworth:2002tt} and SUSY decay
chains \cite{Butterworth:2007ke}, and subsequently reexamined in
Refs.~\cite{Kribs:2009yh,Plehn:2009rk}. Recent studies at the
LHC~\cite{ATLAS-CONF-2011-073, ATL-PHYS-PUB-2009-088,
  CMS-PAS-JME-10-013} have also shown this approach to be promising.

The study in Ref.~\cite{Butterworth:2008iy} was carried out using the
(\textsf{FORTRAN}) \textsf{HERWIG 6.510} event generator~\cite{Corcella:2000bw,Corcella:2002jc}
together with the simulation of the underlying event using \textsf{JIMMY 4.31}~\cite{Butterworth:1996zw}.
In order to allow the inclusion of new theoretical developments and  improvements
in non-perturbative modelling a new simulation based on the same physics philosophy
\textsf{Herwig++}, currently version \textsf{2.6}~\cite{Bahr:2008pv,Arnold:2012fq}, is now preferred 
for the simulation of hadron--hadron collisions.

\textsf{Herwig++} includes both an improved theoretical description
of perturbative QCD radiation, in particular for radiation from heavy quarks,
such as bottom, together with improved non-perturbative modeling, especially
of multiple parton--parton scattering and the underlying event.
In \textsf{FORTRAN} \textsf{HERWIG} a crude implementation of the 
dead-cone effect~\cite{Marchesini:1989yk} meant that there was no radiation
from heavy quarks for evolution scales below the quark mass, rather than a smooth suppression of soft collinear radiation.
In \textsf{Herwig++} an improved choice of evolution variable~\cite{Gieseke:2003rz} allows
evolution down to zero transverse momentum
for radiation from heavy particles and reproduces the correct soft limit.
There have also been significant developments of the multiple-parton scattering model
of the underlying event~\cite{Bahr:2008dy,Bahr:2008wk}, including colour reconnections~\cite{Gieseke:2012ft}
and tuning to LHC data~\cite{Gieseke:2011xy}.

The background to jet substructure searches for the Higgs boson comes from
QCD jets which mimic the decay of a boosted heavy particle. Although
\textsf{Herwig++} has performed well in some early studies of jet
substructure~\cite{CMS-PAS-JME-10-013,ATLAS:2012am,ATLASTEMP}, it is important
that we understand the uncertainties in our modelling of the background
jets which lie at the tail of the jet mass distribution.

 In addition we improve
the simulation of Higgs boson decay by implementing the next-to-leading-order~(NLO)
corrections to 
Higgs boson decay to heavy quarks in the POWHEG~\cite{Nason:2004rx,Frixione:2007vw} formalism.

In the next section we present our approach for the tuning of the parameters,
which effect QCD radiation and hadronization, in \textsf{Herwig++} together
with the results of our new tune. We then recap the key features of the
Butterworth, Davison, Rubin and Salam (BDRS) jet substructure technique of
Ref.\,\cite{Butterworth:2008iy}. This
is followed by our results using both the leading and
next-to-leading-order matrix elements in \textsf{Herwig++} with
implementation of the next-to-leading-order Higgs boson decays and our estimate
on the uncertainties.

\section{Tuning \textsf{Herwig++}}

Any jet substructure analysis is sensitive to
changes in the simulation of initial- and final-state radiation, and hadronization.
In particular the non-perturbative nature of the phenomenological
hadronization model means there are a number of parameters which are
tuned to experimental results.
\textsf{Herwig++} uses an improved angular-ordered parton shower
algorithm~\cite{Gieseke:2003rz,Bahr:2008pv} to describe perturbative QCD
radiation together with a cluster hadronization model~\cite{Webber:1983if,Bahr:2008pv}. 

The \textsf{Herwig++} cluster model is based on the concept of
preconfinement \cite{Amati:1979fg}. At the end of the parton-shower
evolution all gluons are non-perturbatively split into quark-antiquark pairs.
All the partons can then be formed into colour-singlet clusters
which are assumed to be hadron precursors and 
decay according to phase space into the observed
hadrons. There is a small fraction of heavy clusters for which this
is not a reasonable approximation which are therefore first fissioned
into lighter clusters. The main advantage of this model, when
coupled with the angular-ordered parton shower is that it has fewer
parameters than the string model as 
implemented in the \textsf{{\sc{Pythia}}}~\cite{Sjostrand:2006za} event generator
yet still gives a reasonable description of collider observables
\cite{Buckley:2011ms}.

To tune \textsf{Herwig++}, and investigate the dependency of observables on the 
shower and hadronization parameters, the \textsf{Professor} Monte
Carlo tuning system~\cite{Buckley:2009bj} was used. \textsf{Professor}
uses the \textsf{Rivet} analysis framework~\cite{Buckley:2010ar} and 
a number of simulated event samples, with different Monte Carlo parameters, to
parameterise the dependence of each observable\footnote{Normally this is
either an observation such as a multiplicity or a bin in a measured distribution.}
used in the tuning on the parameters of the Monte Carlo event generator.
A  heuristic chi-squared function
\begin{equation}
\chi^{\prime\,2}(\boldsymbol{p}) = 
\sum_{\mathcal{O}}w_{\mathcal{O}}\sum_{b\in\mathcal{O}} 
\frac{\left(f^b(\boldsymbol{p})-\mathcal{R}_b\right)^2}{\Delta^2_b},
\end{equation}
is constructed where $\boldsymbol{p}$ is the set of parameters being tuned, $\mathcal{O}$
 are the observables used each with weight
$w_\mathcal{O}$, $b$ are the different bins in each observable distribution
 with associated experimental measurement $\mathcal{R}_b$,
error $\Delta_b$ and Monte Carlo prediction $f^b(\boldsymbol{p})$.
Weighting of those observables for which a good description of the 
experimental result is important is used in most cases.
The parameterisation of the event generator response, $f(\boldsymbol{p})$, is then
used to minimize the $\chi^{\prime\,2}$ and find the optimum parameter values.

There are ten main free parameters which affect the
shower and hadronization in \textsf{Herwig++}. 
These are shown in
Table~\ref{tab:defaultParams} along with their default values and
allowed ranges.

The gluon mass, \GluonMass,  is required to 
allow the non-perturbative decay of gluons into $q\bar q$ pairs and  controls
the energy release in this process. 
\PSplitLight, \ClPowLight\ and  \ClMaxLight\ control the mass distributions
of the clusters produced during the fission of heavy clusters. \ClSmrLight\ controls
the smearing of the direction of hadrons containing a (anti)quark from the 
perturbative evolution about the direction of the (anti)quark. 
\AlphaMZ\ is strong coupling at the $Z^0$ boson mass and controls
the amount of QCD radiation in the parton shower, while \Qmin\ controls
the infrared behaviour of the strong coupling. \pTmin\ is the minimum
allowed transverse momentum in the parton shower and controls the amount
of radiation and the scale at which the perturbative evolution terminates.
\PwtDIquark\ and \PwtSquark\ are the probabilities of selecting
a diquark-antidiquark or $s\bar s$ quark pair from the vacuum during
cluster splitting, and affect the production of baryons and
strange hadrons respectively.

 Previous experience of tuning \textsf{Herwig++} has
found that \Qmin, \GluonMass, \ClSmrLight \,and \ClPowLight \, to be
flat, and so it was chosen to fix these at their default values~\cite{Bahr:2008pv}.

\begin{table}[t]
  \footnotesize
  \begin{center}
    \begin{tabular}{|l| c| c| c| c|}
      \hline
      Parameter & Default Value & Allowed Range & Scanned Range & Optimum Value \\
      \hline \hline 
      \Qmin & 0.935 & $\geq 0$ & $0.500 - 2.500$ & Fixed at default\\
      \GluonMass & 0.95 & $0 - 1$ & $0.75 - 1.00$ & Fixed at default \\
      \ClSmrLight & 0.78 & $0 - 2$ & $0.30 - 3.00$ & Fixed at default \\
      \ClPowLight & 1.28 & $0 - 10$ & $0.50 - 4.00$ & Fixed at default \\
      \pTmin & 1.00 & $\geq 0$ & $0.50 - 1.50$ & $0.88$ \\
      \AlphaMZ & 0.12 & $\geq 0$ & $0.10 - 0.12$ & $0.11$ \\
      \ClMaxLight & 3.25 & $0 - 10$ & $3.00 - 4.20$ & $3.60$ \\
      \PSplitLight & 1.20 & $0 - 10$ & $1.00 - 2.00$ & $0.90$ \\
      \PwtDIquark & 0.49 & $0 - 10$ & $0.10 - 0.50$ & $0.33$ \\
      \PwtSquark & 0.68 & $0 - 10$ & $0.50 - 0.80$ & $0.64$ \\
      \hline
    \end{tabular}
    \caption{The ten parameters to which the jet substructure is most
    sensitive with their default values, the allowed range of these values in 
    \textsf{Herwig++}, the range scanned over and 
    the new optimum value found from minimizing $\chi^{\prime\,2}$.}
    \label{tab:defaultParams}
  \end{center}
\end{table}

To determine the allowed variation of these parameters 
\textsf{Professor} was used to tune the variables in
Table~\ref{tab:defaultParams} to the observables and weights found in
Appendix \ref{app:observables} in Tables \ref{tab:ALEPH_1996_S3486095},
\ref{tab:DELPHI_1996_S3430090}, \ref{tab:PDG_HADRON_MULTIPLICITIES}
and \ref{tab:DELPHI_2002_069_CONF_603}. The dependence of $\chi^{\prime2}$ on the 
various parameters, about the minimum $\chi^{\prime2}$ value, is then diagonalized.

The variation of the parameters
along the eigenvectors in parameter space obtained corresponding to
a certain change, $\Delta\chi^{\prime\,2}$, in $\chi^{\prime\,2}$ can then be used to predict
the uncertainty in the Monte Carlo predictions for specific observables.

\begin{figure}[t!!]
  \begin{center}
    \subfigure[\pTmin]{
      \includegraphics[width=0.45\textwidth]{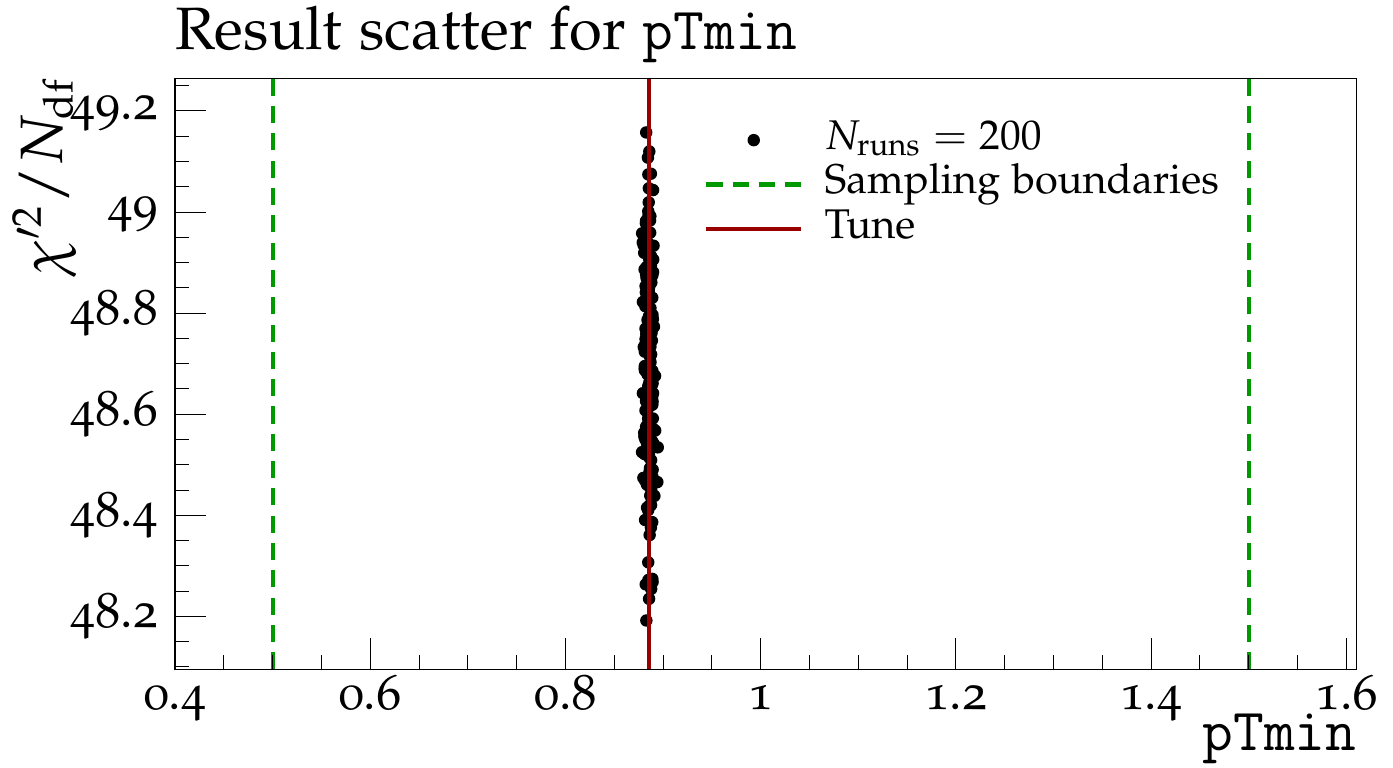}
      \label{subfig:pTminScat}
    }  
    \subfigure[\AlphaMZ]{
      \includegraphics[width=0.45\textwidth]{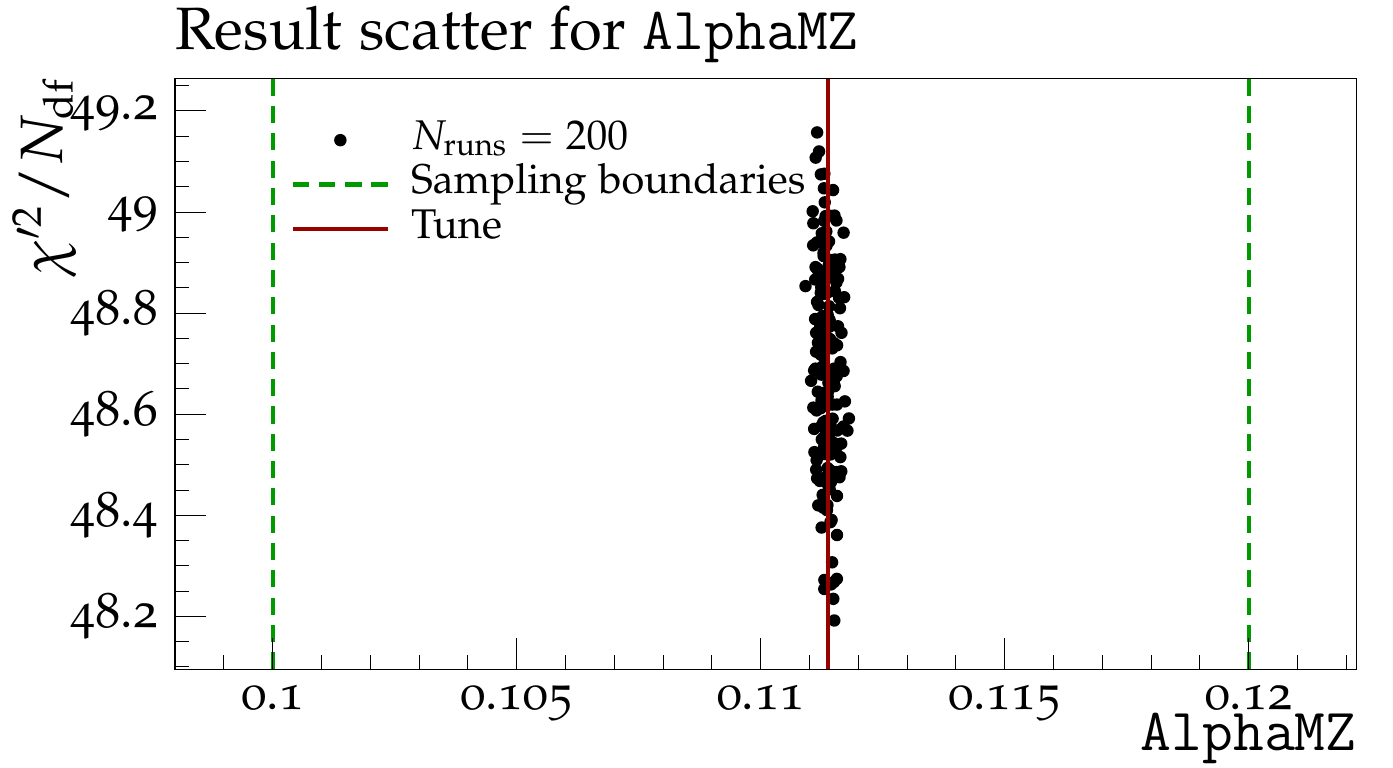}
      \label{subfig:AlphaMZScat}
    }  
    \subfigure[\ClMaxLight]{
      \includegraphics[width=0.45\textwidth]{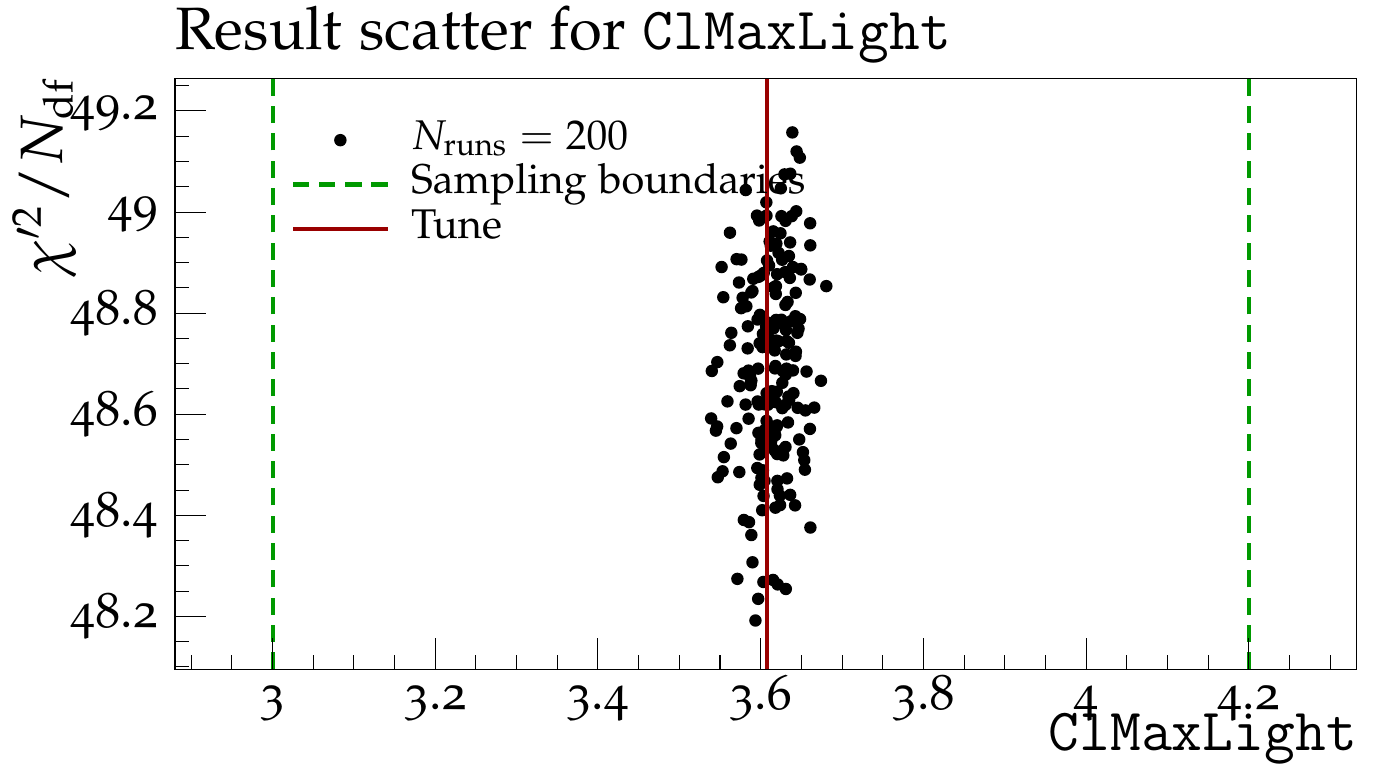}
      \label{subfig:ClMaxLightScat}
    }  
    \subfigure[\PwtDIquark]{
      \includegraphics[width=0.45\textwidth]{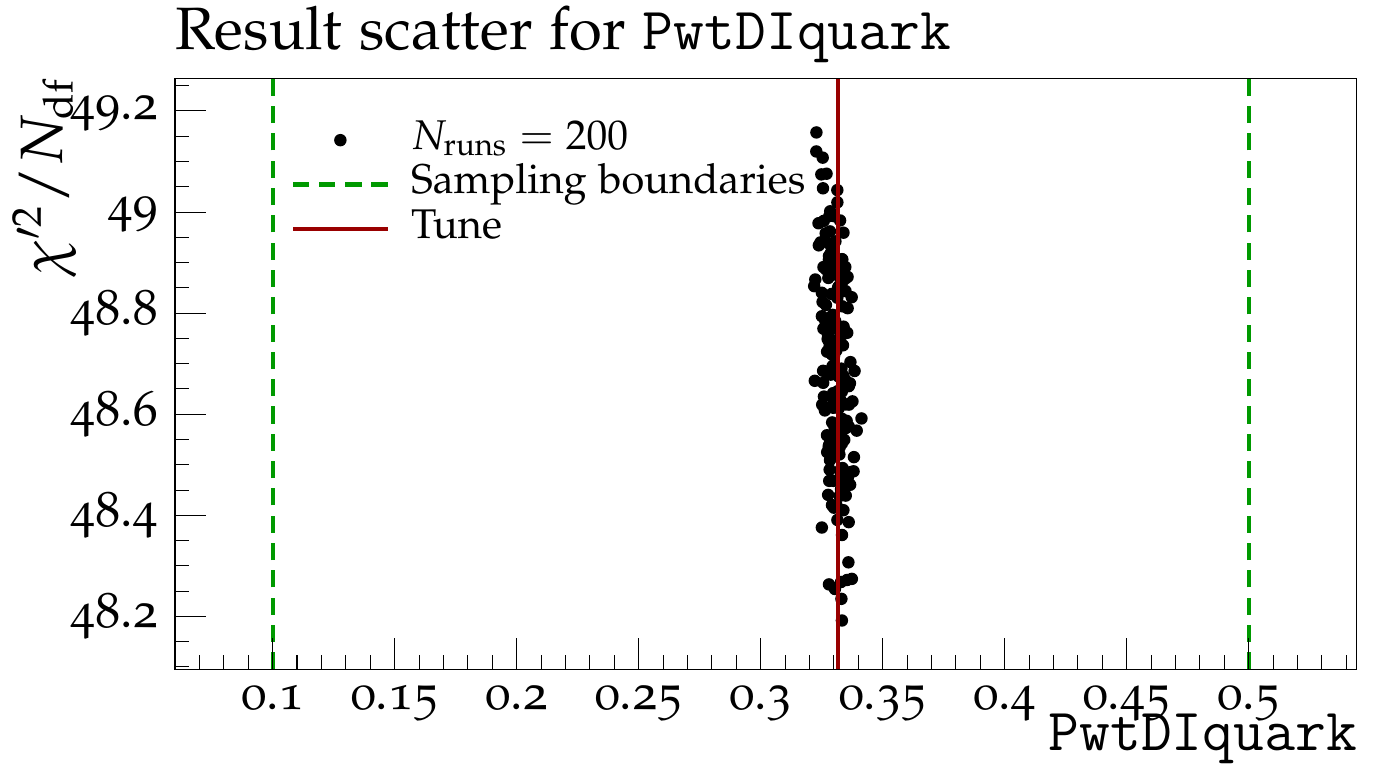}
      \label{subfig:PwtDIquarkScat}
    }  
    \subfigure[\PwtSquark]{
      \includegraphics[width=0.45\textwidth]{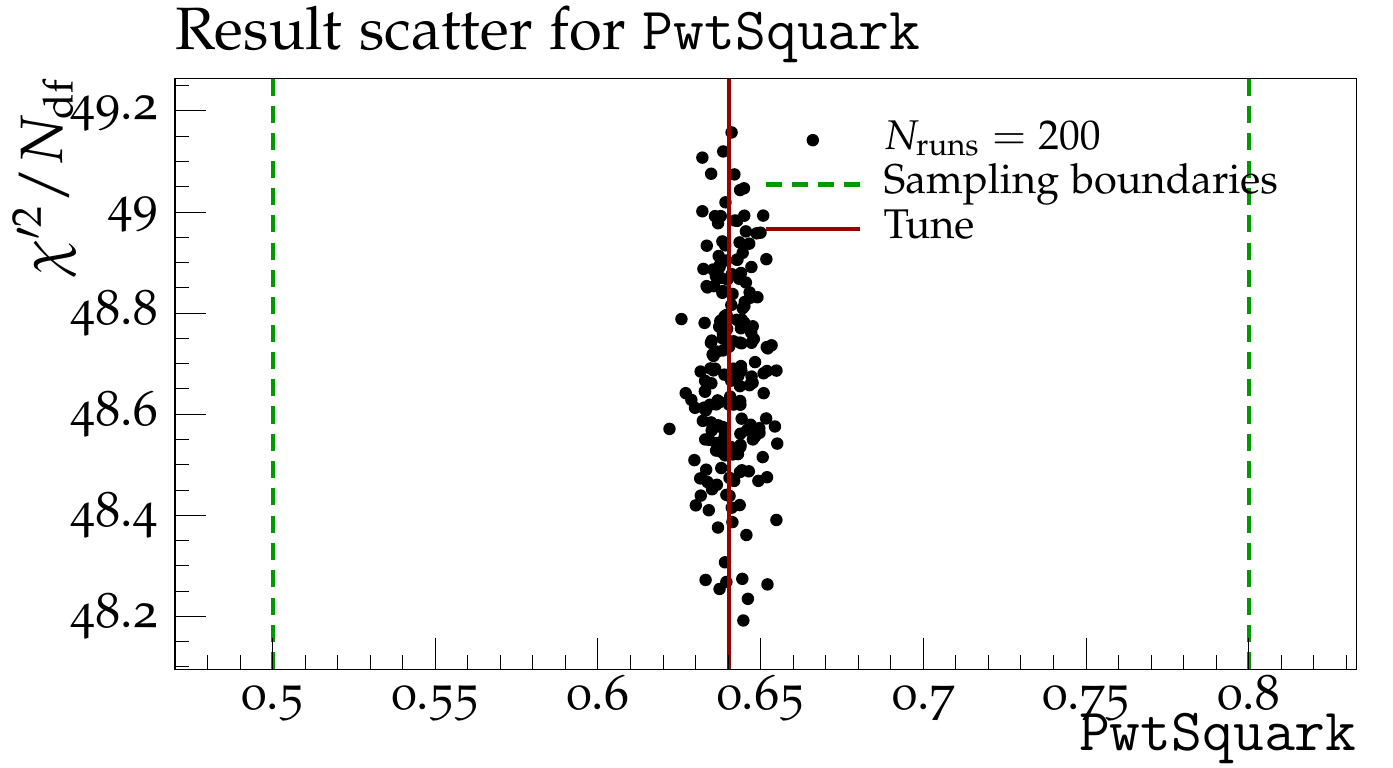}
      \label{subfig:PwtSquarkScat}
    }  
  \end{center}
  \caption{The $\chi^{\prime\,2}/N_{\mathrm{df}}$ distributions for the
    parameters that were varied from their default values
    whilst determining the error tune. The scatter of the results gives
    a representation of the systematics of tuning procedure.}
  \label{fig:ParamsXDist}
\end{figure}

In theory, if the $\chi^{\prime\,2}$ measure for the parameterised generator
response is actually distributed as a true $\chi^2$, then a change in
the goodness of fit of one will correspond to a one sigma deviation from the
minima, {\it i.e.} the best tune. In practice, even the best tune does not
fit the data ideally and nor is the $\chi^{\prime\,2}$ measure actually
distributed according to a true $\chi^2$ distribution. This means that one cannot
just use \textsf{Professor} to vary the parameters about the minima to a given
deviation in the $\chi^{\prime\,2}$ measure without using some subjective opinion on
the quality of the results.

We simulated one thousand event samples with different randomly selected values
of the parameters we were tuning. Six hundred of these were used to interpolate
the generator response. All the event samples were used to select two hundred
samples randomly two hundred times in order to assess the systematics of the 
interpolation and tuning
procedure. A cubic interpolation of the generator response 
was used as this has been shown to
give a good description of the Monte Carlo behaviour in the region of
best generator response \cite{Buckley:2009bj}. 
 The parameters were
varied between values shown in Table~\ref{tab:defaultParams}. The
quality of the interpolation was checked by comparing the
$\chi^{\prime\,2}/N_{\mathrm{df}}$, where $N_{\mathrm{df}}$ is the number of observable bins used in the tune,
in the allowed parameter range on a parameter
by parameter basis for the observables by comparing the interpolation
response with actual generator response at the simulated parameter values. Bad
regions were removed and the interpolation repeated leaving a volume
in the 5-dimensional parameter space where the interpolation worked
well.

Fig.~\ref{fig:ParamsXDist} shows the $\chi^{\prime\,2}/N_{df}$
distributions for two hundred tunes based on two hundred randomly selected event samples
points for the cubic interpolation. The spread of these values gives an idea of the
systematics of the tuning process showing that we have obtained a good fit for
our parameterisation of the generator response.

The line indicates the tune which is
based on a cubic interpolation from six hundred event samples. It is this 
interpolation which was used to vary $\chi^{\prime\,2}$ about the minimum to
assess the uncertainty on the measured distributions.
 During the tune
it was discovered that \PSplitLight \, was relatively insensitive to the 
 observables used in the tune. As such, \PSplitLight \, was fixed at the default value 
of 1.20 during the tune and subsequent $\chi^{\prime\,2}$ variation.

\textsf{Professor} was used to vary $\chi^{\prime\,2}$ about the minimum value, as described
above, determining the allowed range for the parameters. 
As five parameters were eventually varied, there are 10
new sample points - one for each of the parameters and one ``+'' and
one ``-'' along each eigenvector direction in parameter space.

\begin{figure}[h!!!]
  \begin{center}
    \subfigure[Out-of-plane $p_T$]{
      \includegraphics[width=0.45\textwidth]{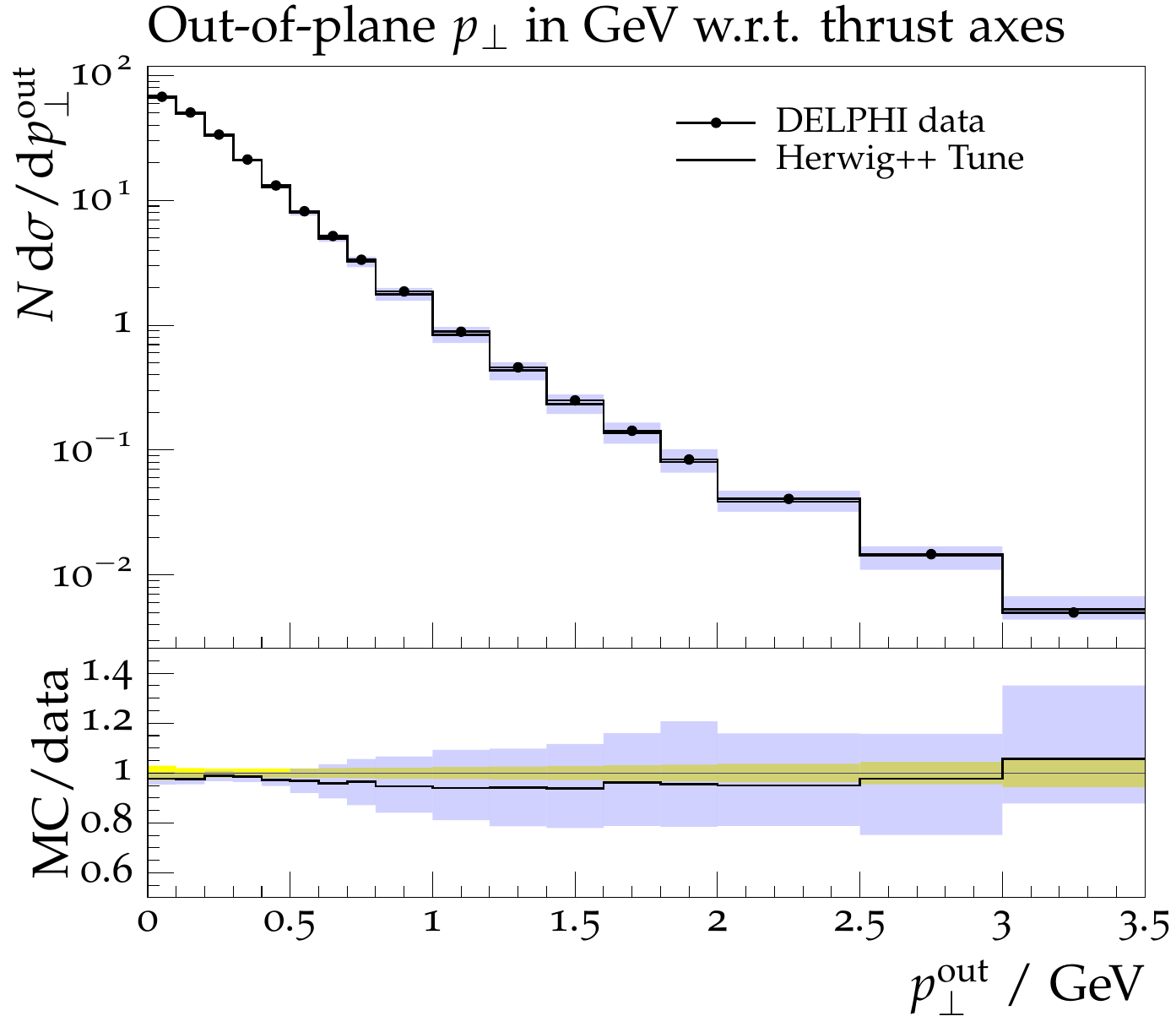}
      \label{subfig:delphid02-5}
    }  
    \subfigure[1-Thrust]{
      \includegraphics[width=0.45\textwidth]{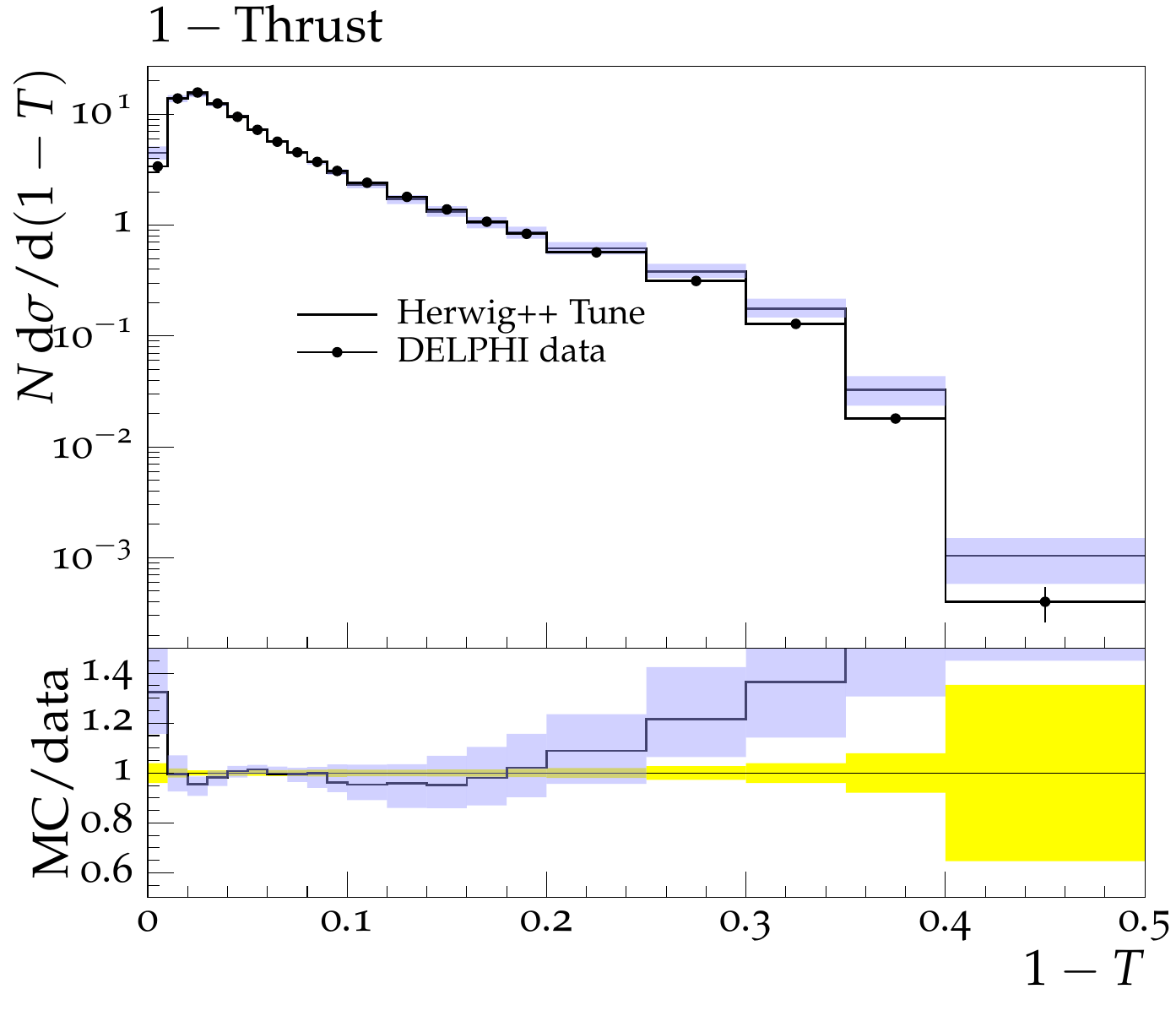}
      \label{subfig:delphid11-5}
    }
  \end{center}
  \caption{Results from the DELPHI~\protect\cite{Abreu:1996na} analysis of out-of-plane $p_T$ with-respect-to the thrust axis and 1-thrust
            showing the new tune and the envelopes corresponding to a change in $\Delta \chi^{\prime\,2}/N_{\mathrm{df}} =5$.}
  \label{fig:delphi5}
  \begin{center}

    \subfigure[Out-of-plane $p_T$]{
      \includegraphics[width=0.45\textwidth]{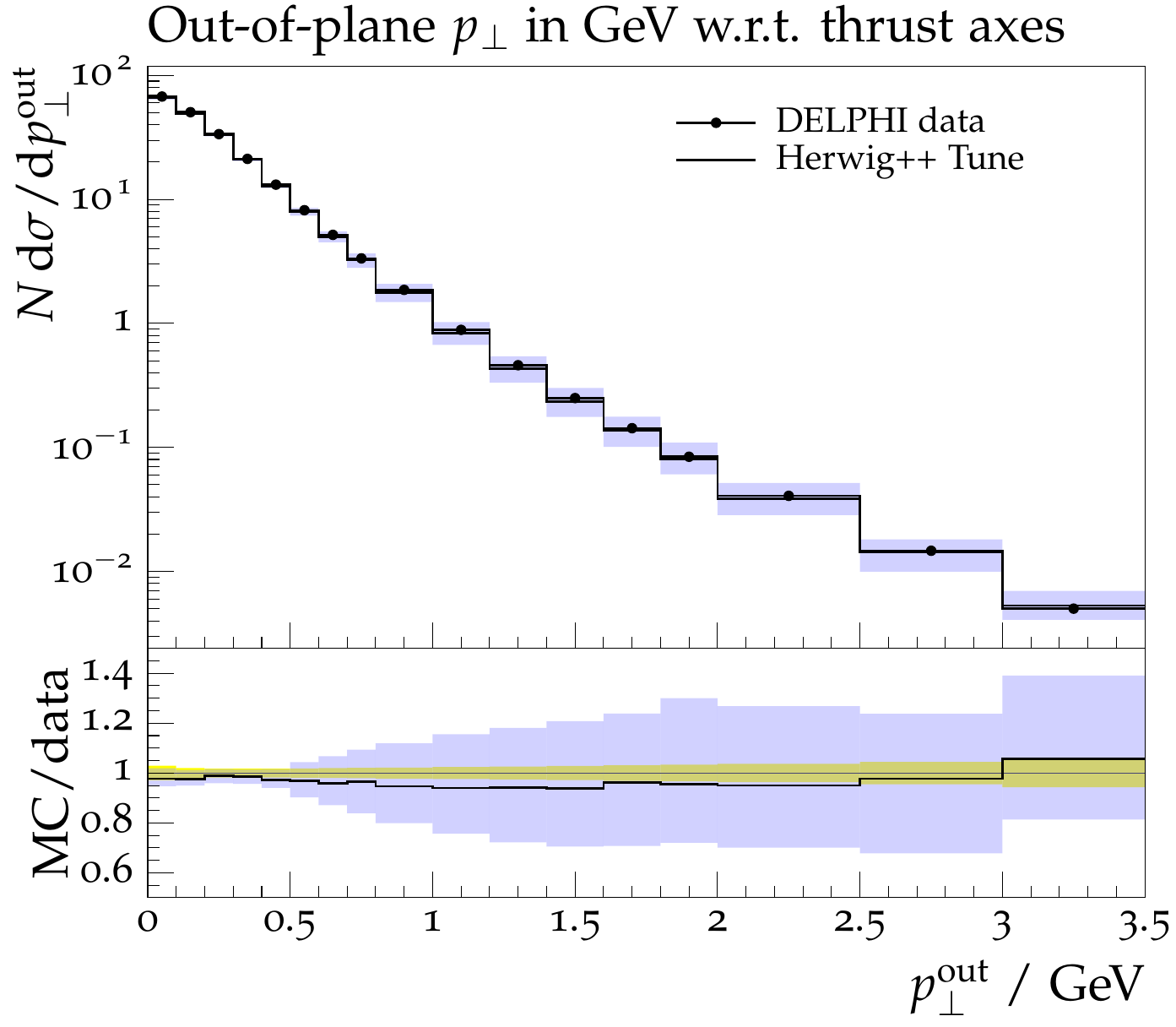}
      \label{ssubfig:delphid02-10}
    }
    \subfigure[1-Thrust]{
      \includegraphics[width=0.45\textwidth]{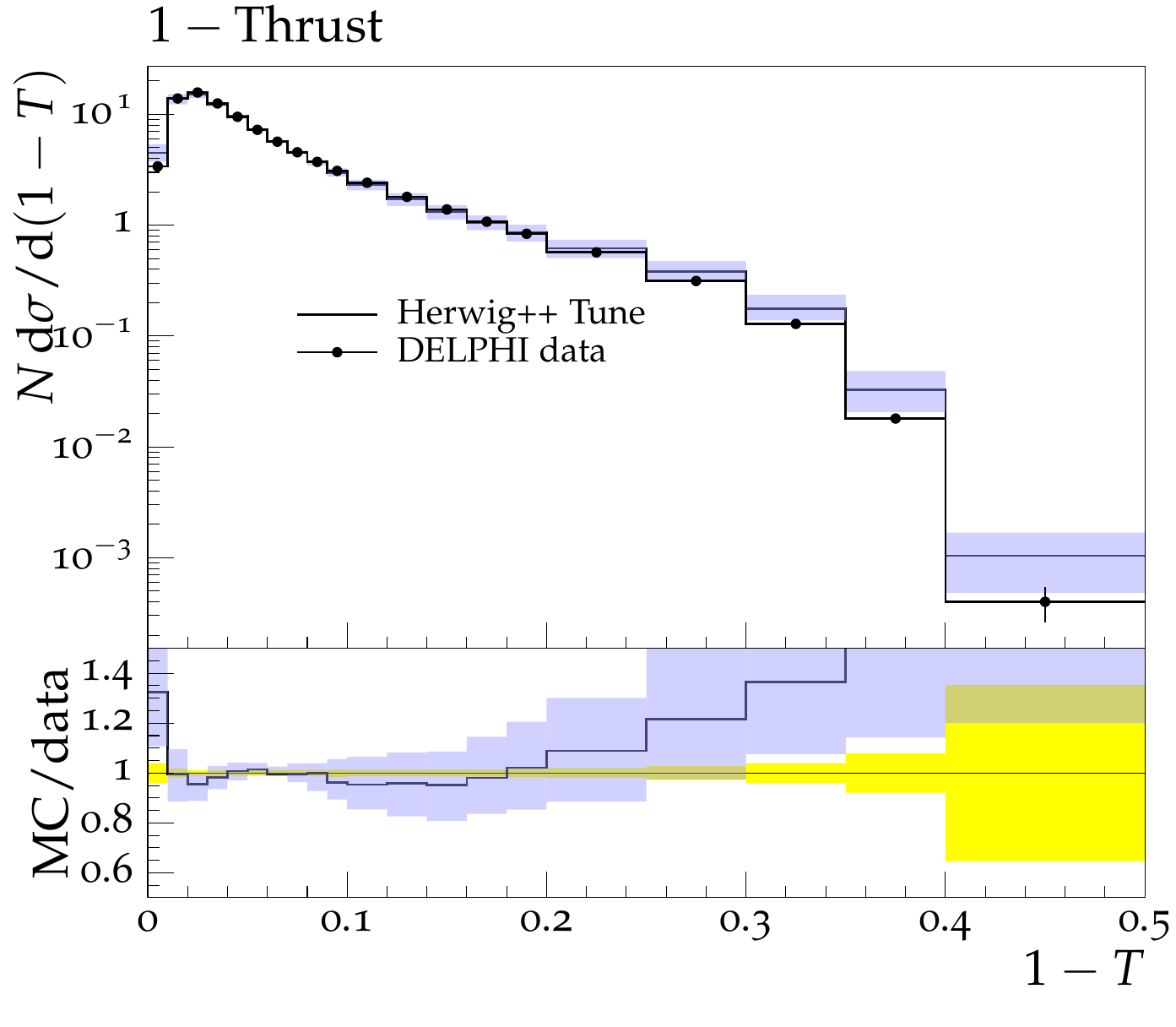}
      \label{subfig:delphid11-10}
    }
  \end{center}
  \caption{Results from the DELPHI~\protect\cite{Abreu:1996na}  analysis of out of plane $p_T$ with-respect-to the thrust axis and 1-thrust showing the new tune and the envelopes corresponding to a change in $\Delta \chi^{\prime\,2}/N_{\mathrm{df}} =10$.}
  \label{fig:delphi10}
\end{figure}

Various changes in $\chi^{\prime\,2}$ were examined. A variation of $\Delta \chi^{\prime\,2}/N_{\mathrm{df}} = 5$ keeps the LEP data within reasonable limits while a variation of
$\Delta \chi^{\prime\,2}/N_{\mathrm{df}} = 10$ is too large.  The values for both $\Delta \chi^{\prime\,2}/N_{\mathrm{df}}
= 5$ and $\Delta \chi^{\prime\,2}/N_{\mathrm{df}} = 10$ are shown in are shown in
Tables~\ref{tab:errorParams5} and Tables~\ref{tab:errorParams10}
respectively.

\begin{table}[t]
  \footnotesize
  \begin{center}
    \begin{tabular}{|l| c| c|| c| c|| c| c|| c| c|| c| c|}
      \hline
      \multirow{3}{*}{Parameter} & \multicolumn{10}{|c|}{Direction} \\  \cline{2-11}
      & \multicolumn{2}{|c||}{1} & \multicolumn{2}{|c||}{2}& \multicolumn{2}{|c||}{3}& \multicolumn{2}{|c||}{4}& \multicolumn{2}{|c|}{5}  \\ 
      \cline{2-11} & + & - & + & - & + & - & + & - & + & - \\  
      \hline \hline
      \pTmin & 0.88 & 0.88 & 0.88 & 0.88 & 0.84 & 0.93 & 0.87 & 0.90 & 0.89 & 0.87 \\ 
      \AlphaMZ & 0.11 & 0.11 & 0.10 & 0.12 & 0.12 & 0.11 & 0.12 & 0.11 & 0.12 & 0.11 \\ 
      \ClMaxLight & 3.61 & 3.61 & 3.61 & 3.61 & 3.60 & 3.62 & 3.66 & 3.55 & 3.54 & 3.67 \\ 
      \PwtDIquark & 0.46 & 0.23 & 0.33 & 0.33 & 0.33 & 0.33 & 0.33 & 0.33 & 0.33 & 0.33 \\ 
      \PwtSquark & 0.64 & 0.64 & 0.64 & 0.64 & 0.64 & 0.64 & 0.62 & 0.67 & 0.51 & 0.78 \\ 
      \hline
    \end{tabular}
    \caption{The five directions corresponding to the error tune for a $\Delta \chi^{\prime\,2}/N_{\mathrm{df}} = 5$ 
      and the values the parameters take in each direction.}
    \label{tab:errorParams5}
  \end{center}
%
  \footnotesize
  \begin{center}
    \begin{tabular}{|l| c| c|| c| c|| c| c|| c| c|| c| c|}
      \hline
      \multirow{3}{*}{Parameter} & \multicolumn{10}{|c|}{Direction} \\  \cline{2-11}
      & \multicolumn{2}{|c||}{1} & \multicolumn{2}{|c||}{2}& \multicolumn{2}{|c||}{3}& \multicolumn{2}{|c||}{4}& \multicolumn{2}{|c|}{5}  \\ 
      \cline{2-11} & + & - & + & - & + & - & + & - & + & - \\  
      \hline \hline
      \pTmin & 0.88 & 0.88 & 0.88 & 0.88 & 0.82 & 0.95 & 0.86 & 0.90 & 0.89 & 0.87 \\ 
      \AlphaMZ & 0.11 & 0.11 & 0.10 & 0.12 & 0.12 & 0.10 & 0.12 & 0.10 & 0.12 & 0.11 \\ 
      \ClMaxLight & 3.61 & 3.61 & 3.61 & 3.61 & 3.59 & 3.63 & 3.68 & 3.52 & 3.52 & 3.70 \\ 
      \PwtDIquark & 0.51 & 0.19 & 0.33 & 0.33 & 0.33 & 0.33 & 0.33 & 0.33 & 0.33 & 0.33 \\ 
      \PwtSquark & 0.64 & 0.64 & 0.64 & 0.64 & 0.65 & 0.64 & 0.61 & 0.68 & 0.46 & 0.84 \\ 
      \hline
    \end{tabular}
    \caption{The five directions corresponding to the error tune for a $\Delta \chi^{\prime\,2}/N_{\mathrm{df}} = 10$ 
      and the values the parameters take in each direction.}
    \label{tab:errorParams10}
  \end{center}
\end{table}

\begin{figure}[h!!]
  \begin{center}
    \includegraphics[width=0.65\textwidth]{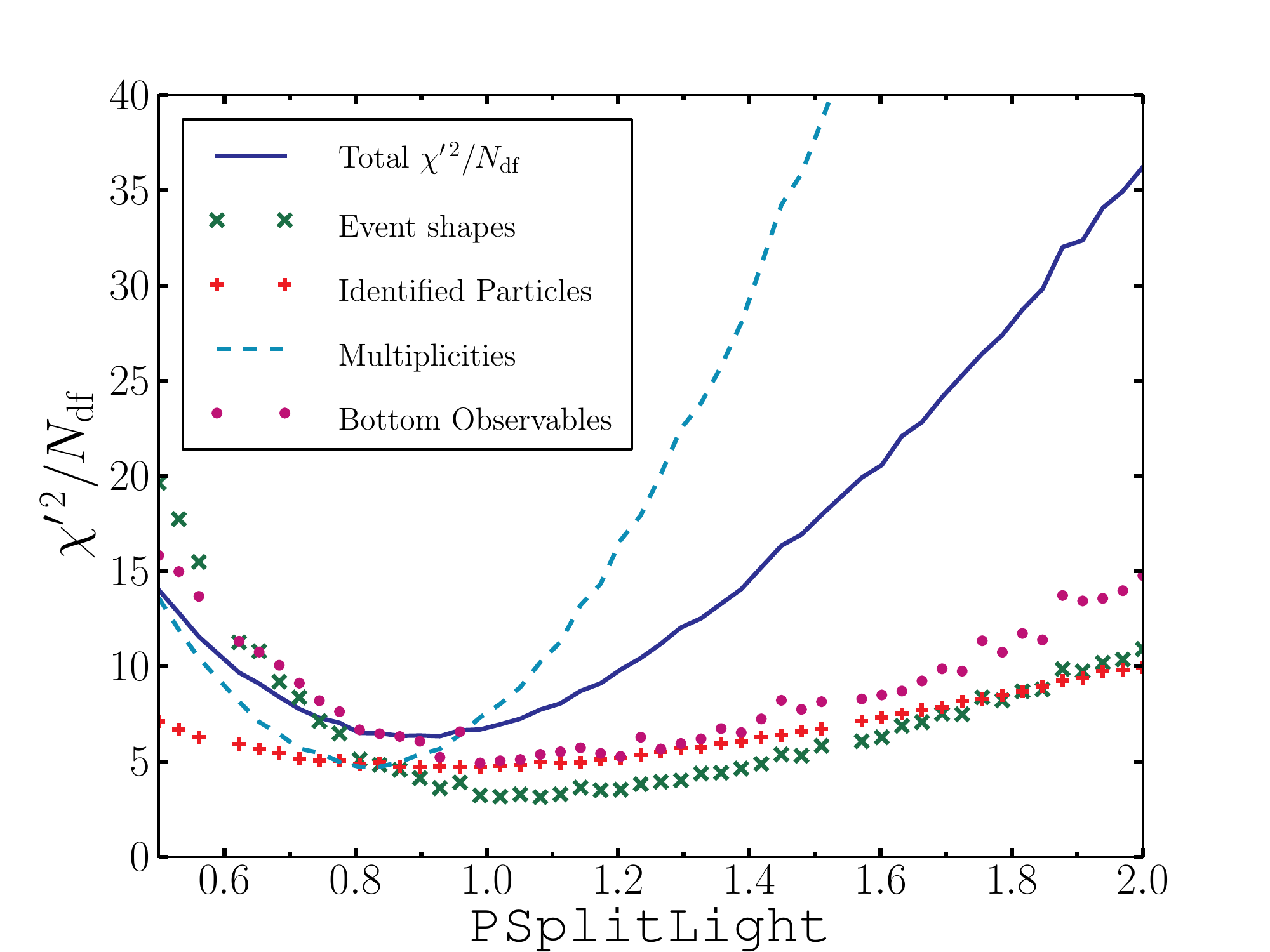}
  \end{center}
  \label{fig:PSplitLight}
  \caption{A scan of \PSplitLight \, using the internal \textsf{Herwig++}
  tuning system with the other parameters fixed at their new tuned value. 
  From the total $\chi^{\prime\,2}/N_{\mathrm{df}}$ we see that a value of $0.90$
  for \PSplitLight \, is favoured at the new tuned parameters driven 
  by the multiplicities.}
\end{figure}

The \textsf{Professor} tune was then compared with the internal
\textsf{Herwig++} tuning procedure \cite{Bahr:2008pv} as not all
analyses that are in the internal \textsf{Herwig++} tuning system are
available in \textsf{Rivet} and subsequently accessible to
\textsf{Professor}. Looking at Fig.~\ref{fig:PSplitLight} it is
found that \PSplitLight \, at a value of $0.90$ is favoured and gives a
significant reduction in the $\chi^{\prime\,2}/N_{\mathrm{df}}$.  It was
therefore decided to use the values obtained from minimisation
procedure, but using the value of $0.90$ for \PSplitLight \, to
maintain a good overall description of the data. The new minima for
the QCD parameters are summarized in the Table~\ref{tab:defaultParams}.
Examples of the new tune and the uncertainty band are shown in
Figs.\,\ref{fig:delphi5}~and~\ref{fig:delphi10} for
the out-of-plane transverse momentum and thrust measured by DELPHI~\cite{Abreu:1996na}.

These error tune values can now be used to predict the uncertainty
from the tuning of the shower parameters on any observable. In the next
section we will present an example of using these tunes to 
estimate the uncertainty on the predictions for searches for the
Higgs boson using the BDRS jet substructure method.

\section{Jet Substructure Boosted Higgs}

\begin{figure}[t!!]
  \begin{center}
    \subfigure[Selection criterion (a)]{
      \includegraphics[width=0.45\textwidth]{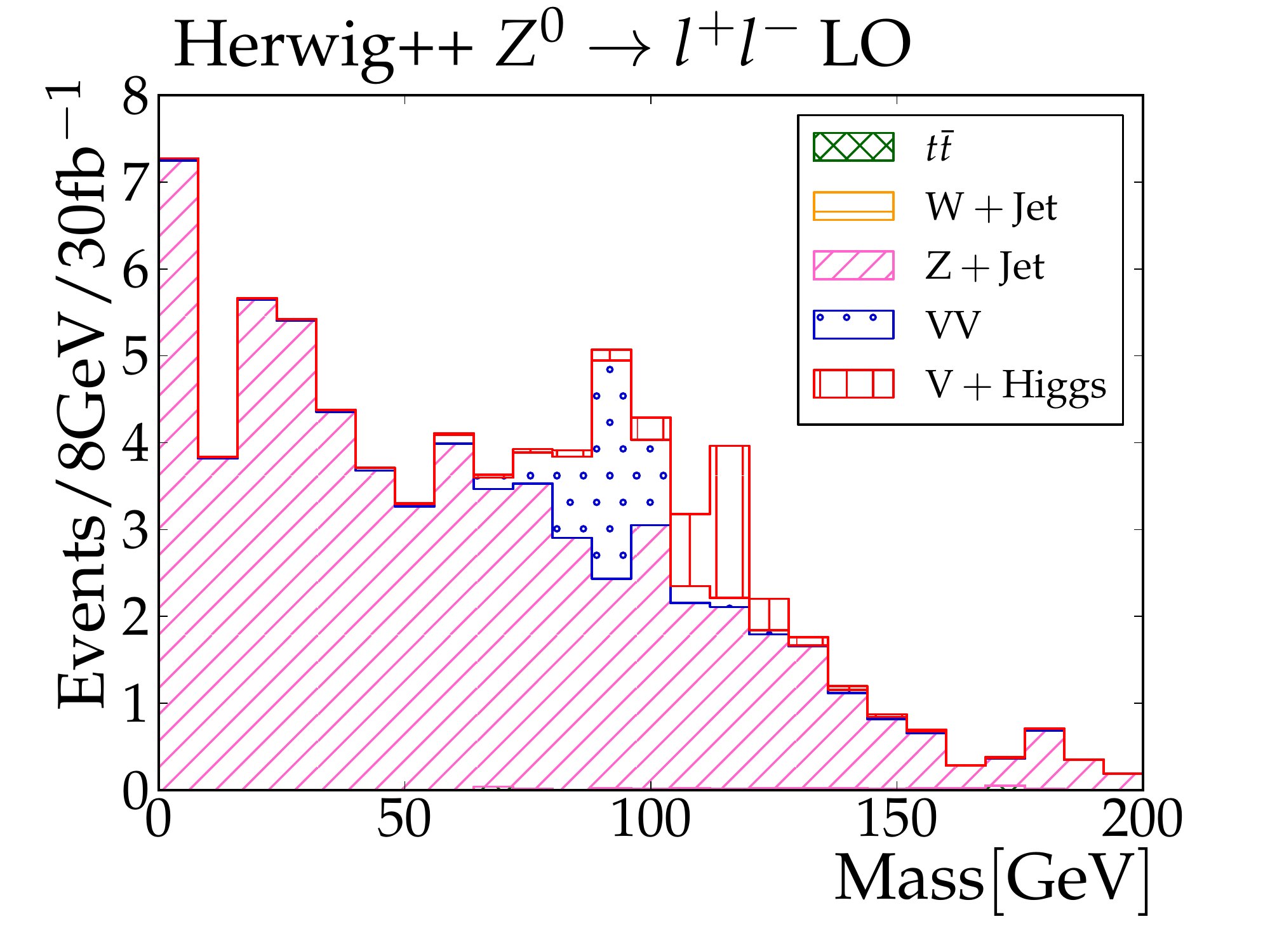}
      \label{subfig:HiggsMassALO}
    }  
    \subfigure[Selection criterion (b)]{
      \includegraphics[width=0.45\textwidth]{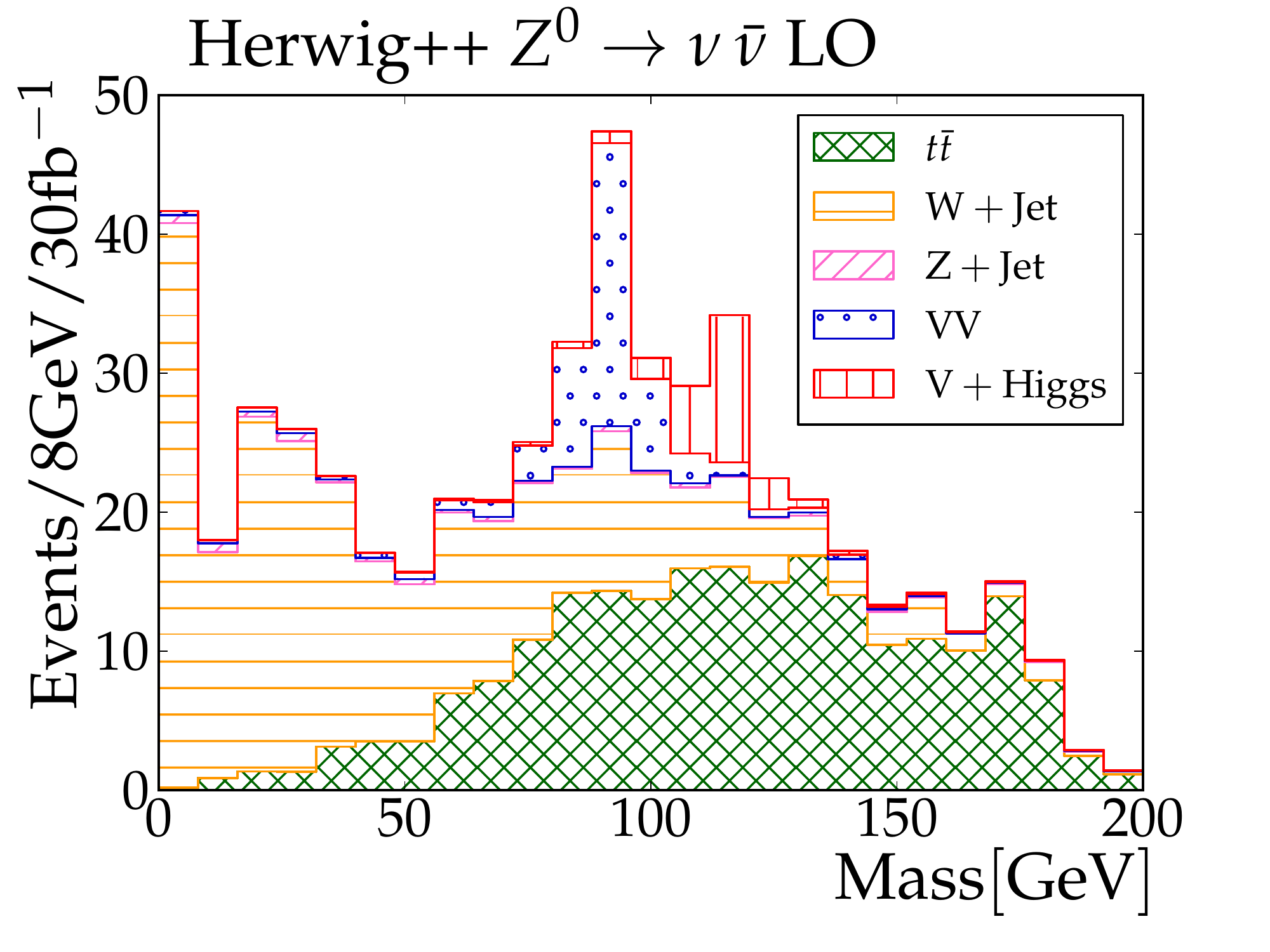}
      \label{subfig:HiggsMassBLO}
    }
    \subfigure[Selection criterion (c)]{
      \includegraphics[width=0.45\textwidth]{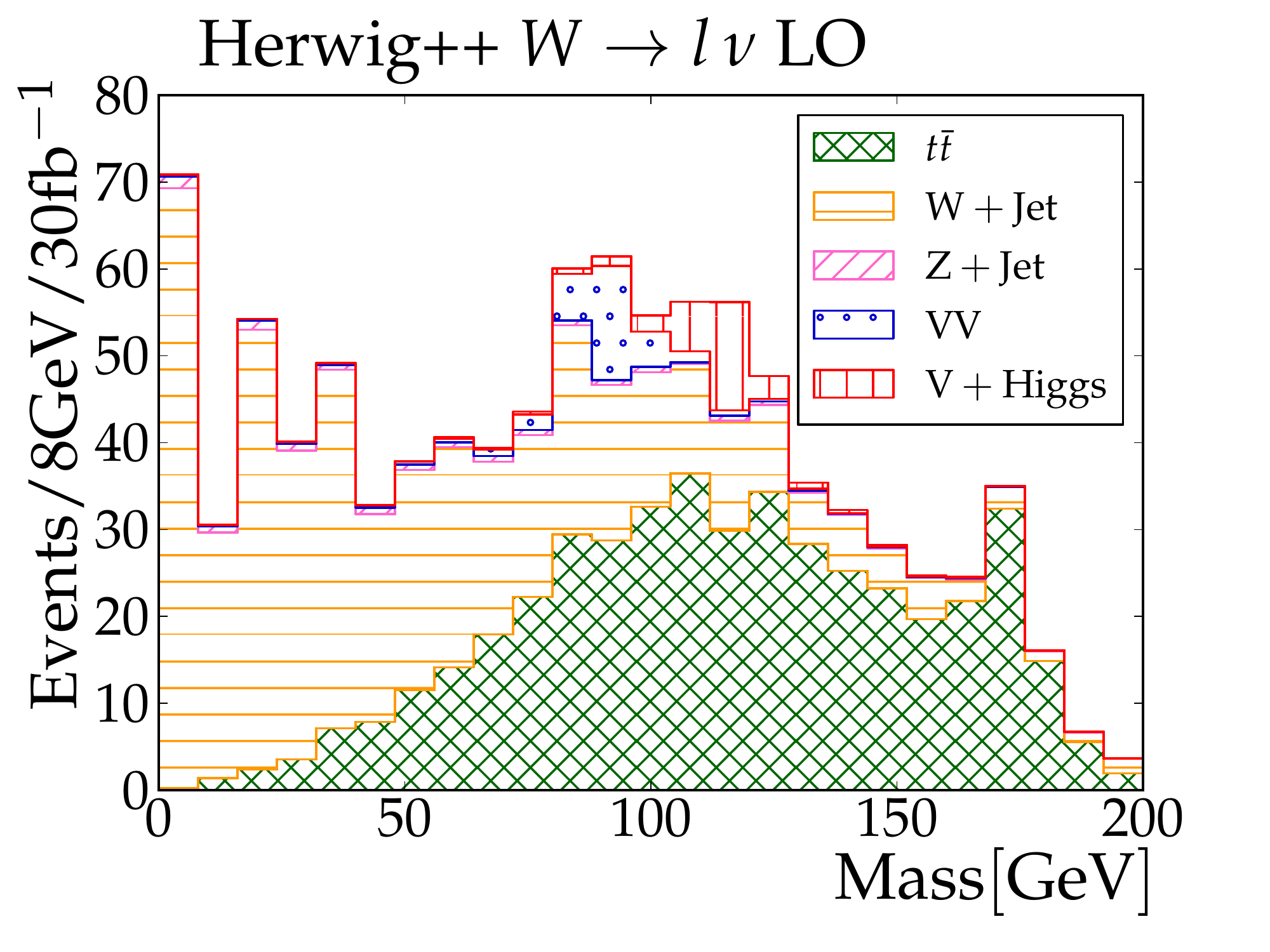}
      \label{subfig:HiggsMassCLO}
    }
    \subfigure[Sum of criteria (a), (b) and (c)]{
      \includegraphics[width=0.45\textwidth]{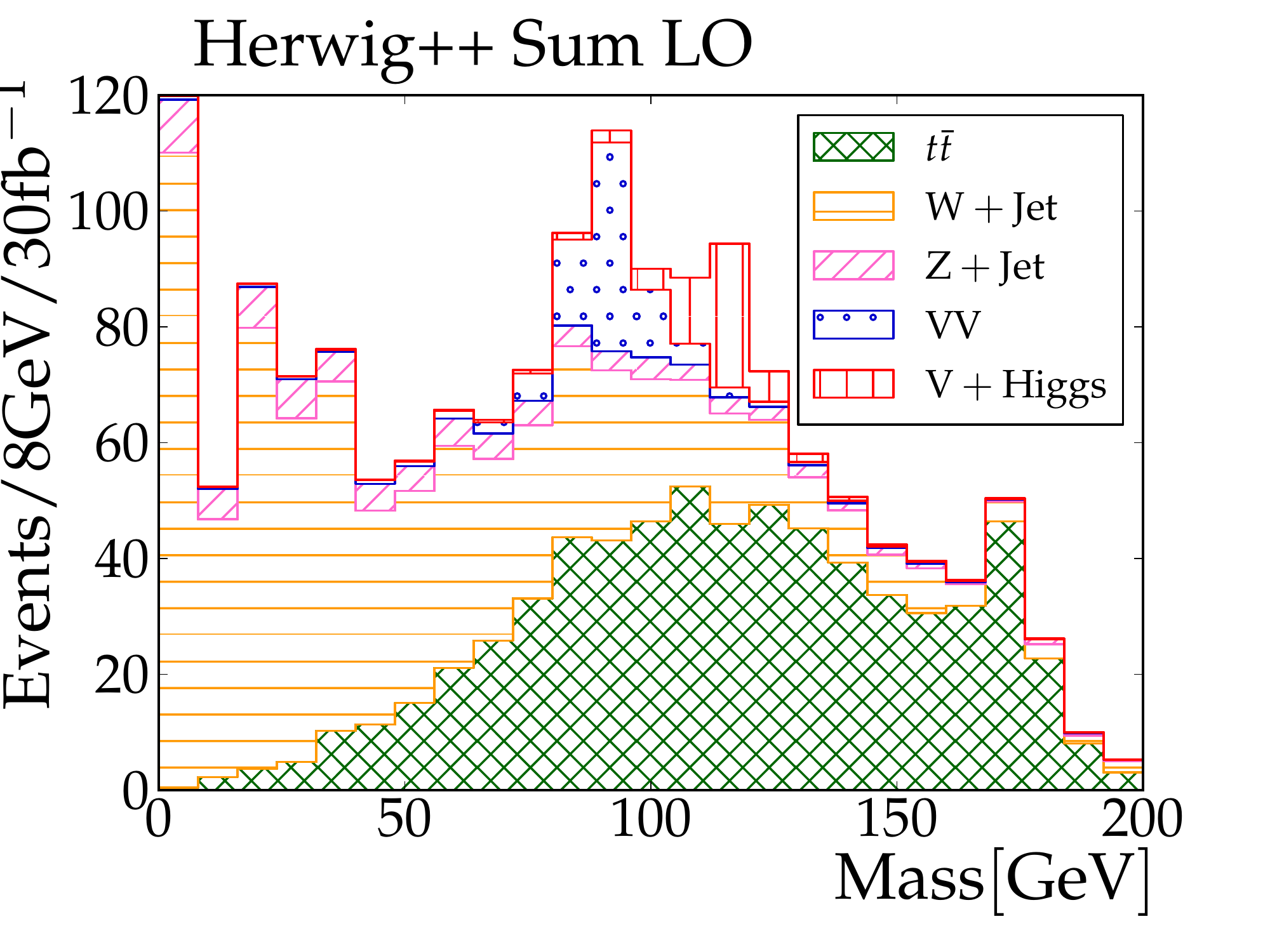}
      \label{subfig:HiggsMassAddedLO}
    }
  \end{center}
  \caption{Results for the reconstructed Higgs boson mass distribution using leading-order matrix elements.
           A SM Higgs boson was assumed with a mass of 115 $\mathrm{GeV}$. In addition to the full result the
           contribution from top quark pair production~($t\bar t$), the production of $W^\pm$~(W+Jet) 
           and $Z^0$~(Z+Jet) bosons in association with a hard jet, vector boson pair production~(VV)
           and the production of a vector boson in association with the Higgs boson~(V+Higgs), are
           shown.}
  \label{fig:loPlots}
\end{figure}
The analysis of Ref.~\cite{Butterworth:2008iy} uses a number of
different channels for the production of the Higgs boson decaying
to $b\bar{b}$ in association with an electroweak gauge boson, {\it i.e.} the
production of $h^0Z^0$ and $h^0W^\pm$. Ref.~\cite{Butterworth:2008iy} uses the fact that the Higgs boson
predominantly decays to $b\,\bar{b}$ in a jet substructure analysis to
extract the signal of a boosted Higgs boson above the various backgrounds.  Their study
found that the Cambridge-Aachen 
algorithm~\cite{Dokshitzer:1997in,Wobisch:2000dk} with radius parameter
$\mathrm{R}=1.2$  gave the best results when combined 
with their jet substructure technique. For our study,
we used the Cambridge-Aachen algorithm as implemented in the \textsf{FastJet}
package \cite{Cacciari:2005hq}.
Three different event selection criteria are used:
\begin{enumerate}[(a)]
  \item a lepton pair with $80 \, \mathrm{GeV} < m_{l^+l^-} < 100 \, \mathrm{GeV}$ and
    $p_T > p_T^{\min}$
   to select events for \mbox{$Z^0\to\ell^+\ell^-$};
  \item missing transverse momentum $\slashed p_T > p_T^{\mathrm{min}}$ to select events
   with \mbox{$Z^0\to\nu\bar\nu$};
  \item missing transverse momentum $\slashed p_T > 30 \, \mathrm{GeV}$ and a lepton with 
    $p_T > 30 \, \mathrm{GeV}$ consistent with the presence of a $\mathrm{W}$ boson with $p_T > p_T^{\min}$ to select events with $W\to\ell\nu$;
    \label{analysis:W}
\end{enumerate}
where $p_T^{\min} = 200 \, \mathrm{GeV}$.

\begin{figure}[t!!]
  \begin{center}
    \subfigure[Selection criterion (a)]{
      \includegraphics[width=0.45\textwidth]{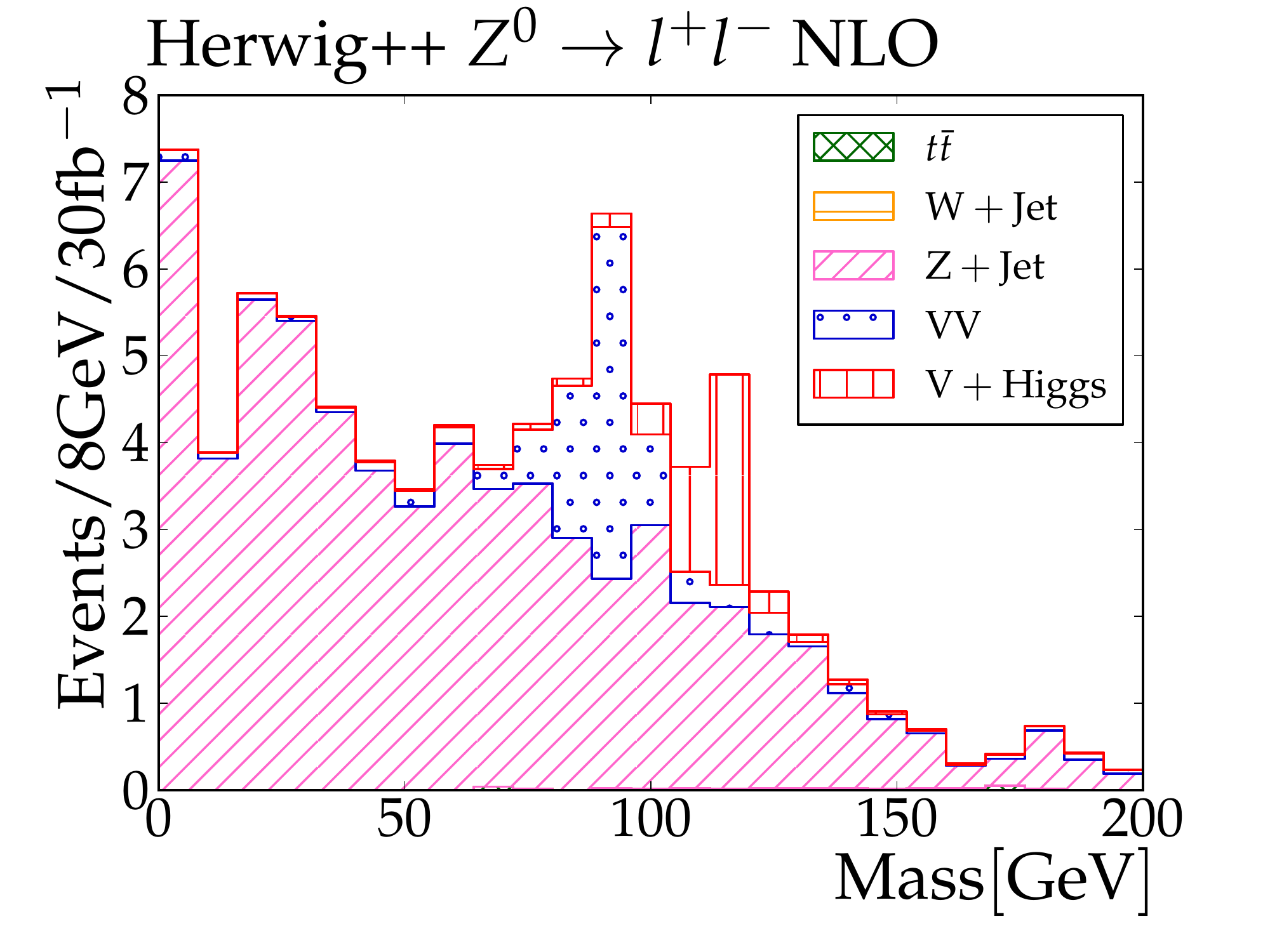}
      \label{subfig:HiggsMassANLO}
    }  
    \subfigure[Selection criterion (b)]{
      \includegraphics[width=0.45\textwidth]{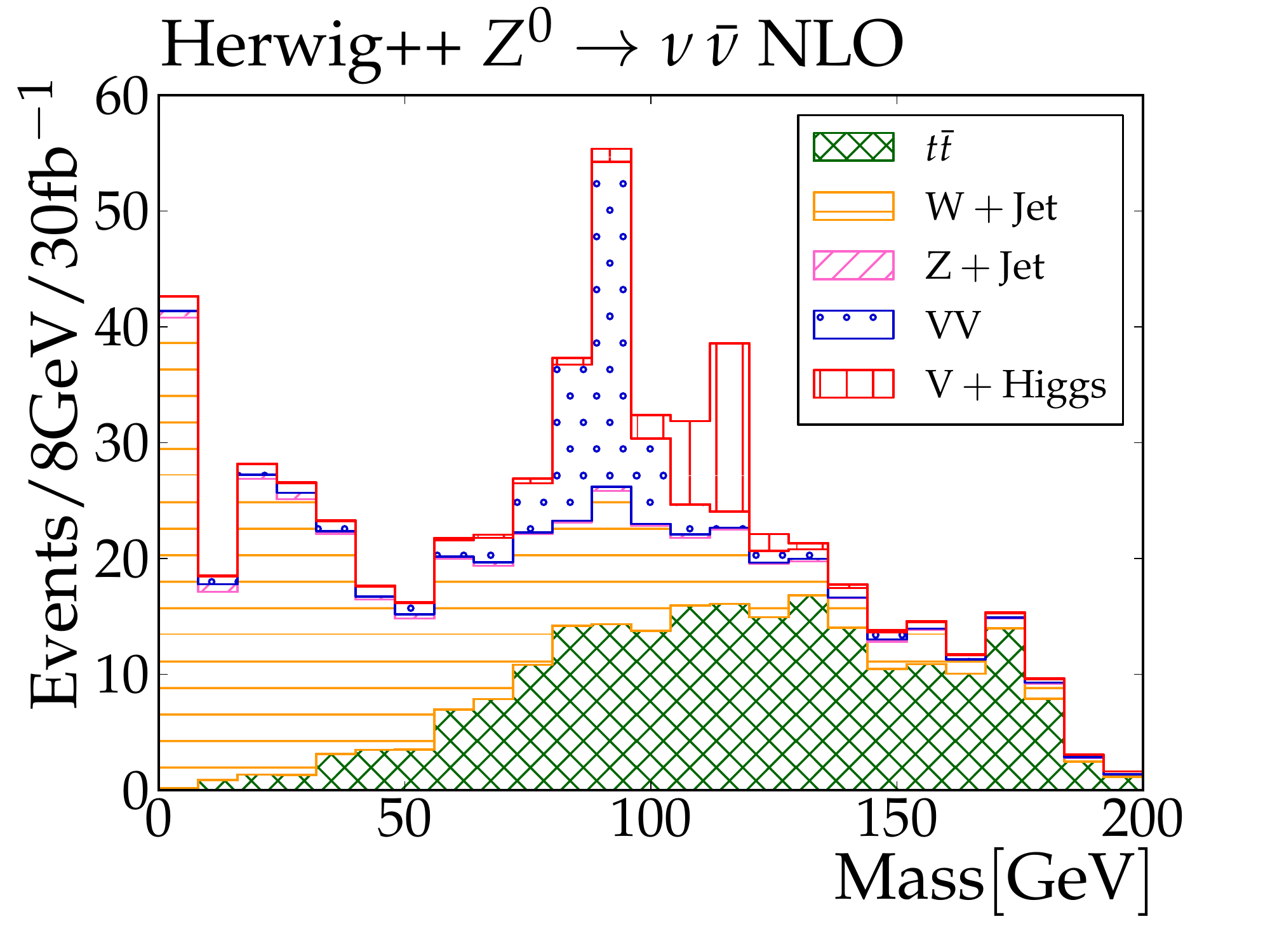}
      \label{subfig:HiggsMassBNLO}
    }
    \subfigure[Selection criterion (c)]{
      \includegraphics[width=0.45\textwidth]{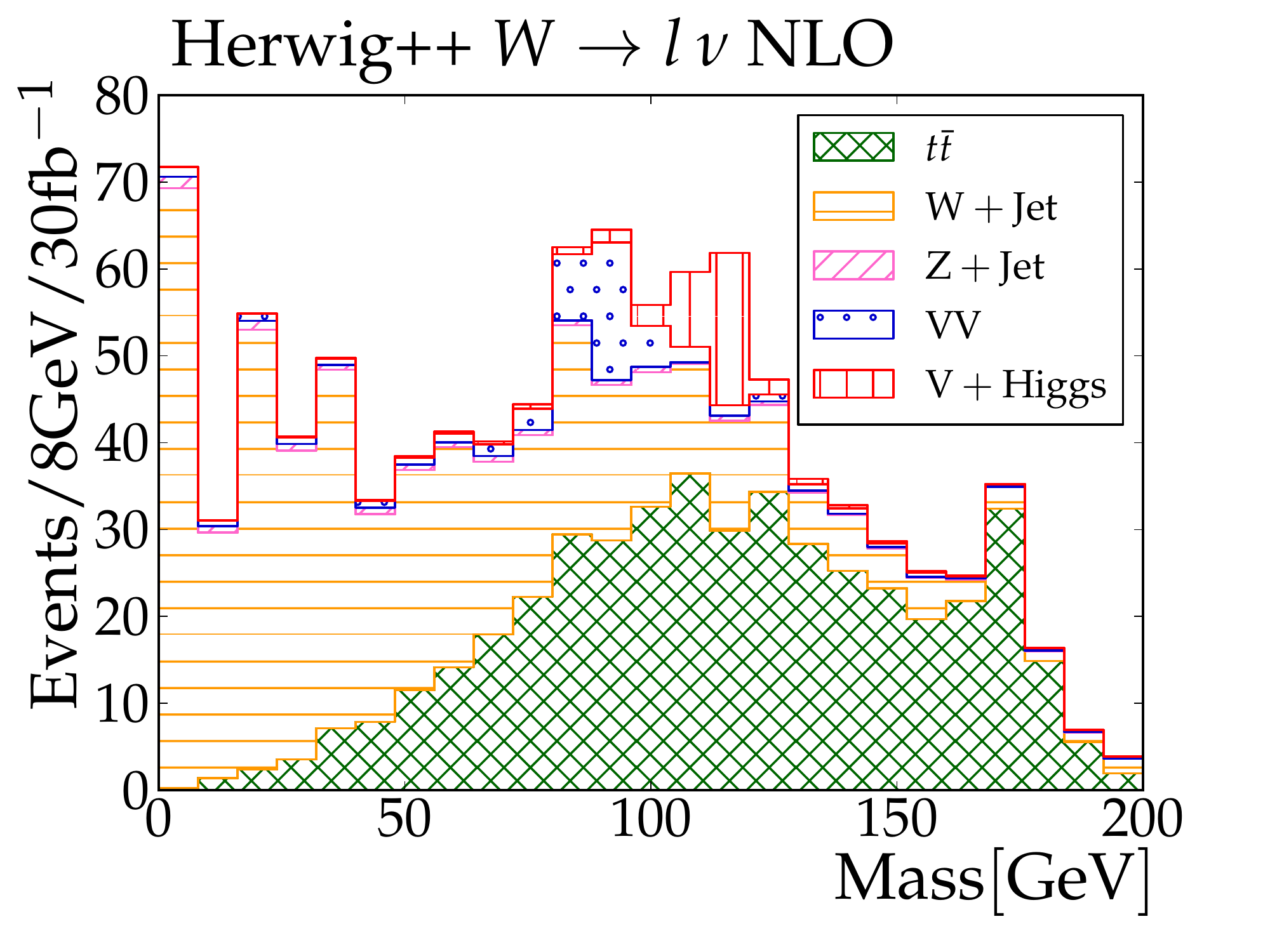}
      \label{subfig:HiggsMassCNLO}
    }
    \subfigure[Sum of criteria (a), (b) and (c)]{
      \includegraphics[width=0.45\textwidth]{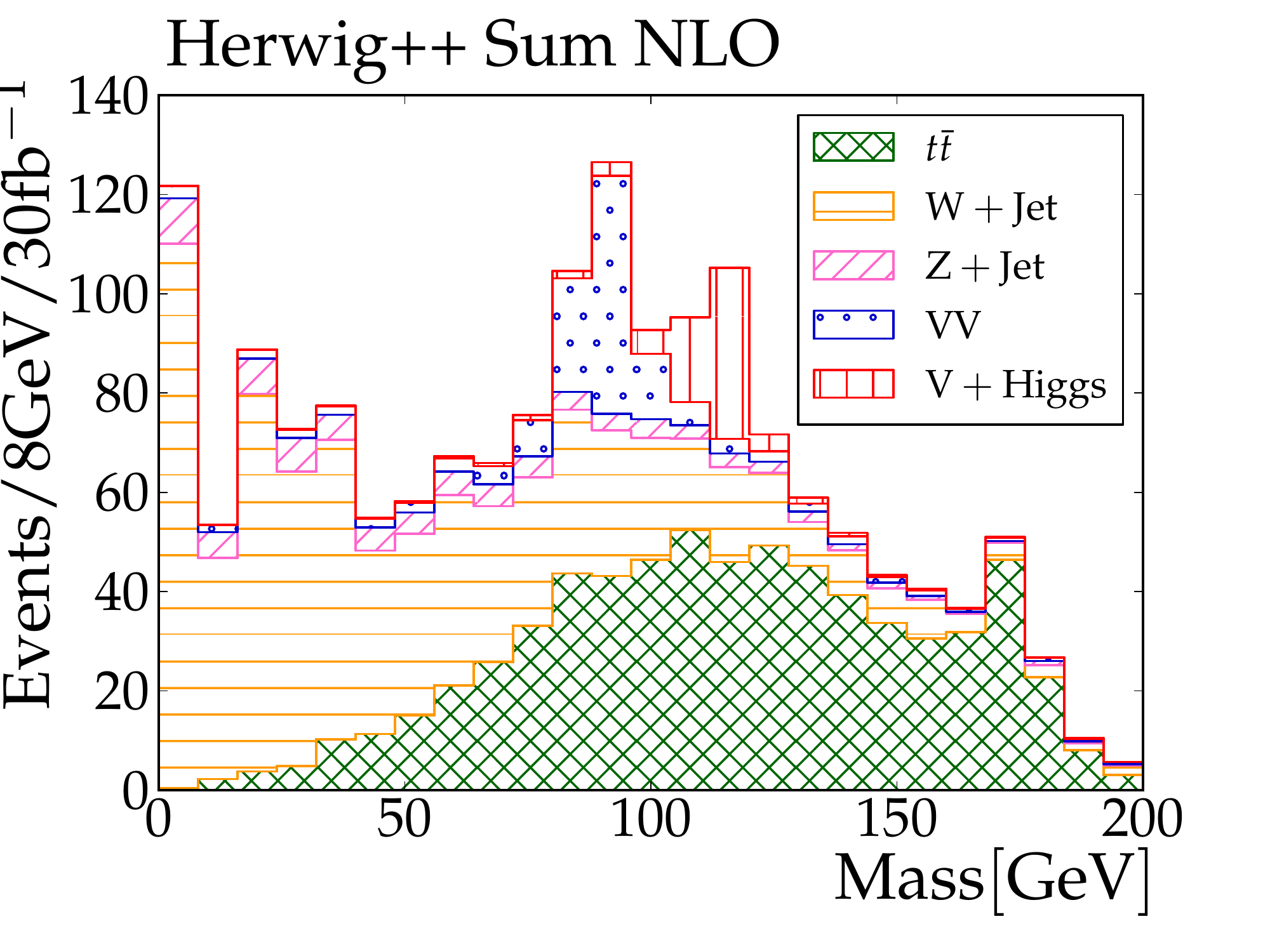}
      \label{subfig:HiggsMassAddedNLO}
    }
  \end{center}
  \caption{Results for the reconstructed Higgs boson mass distribution using
           leading-order matrix elements for  top quark pair production~($t\bar t$), and
           the production of $W^\pm$~(W+Jet) 
           and $Z^0$~(Z+Jet) bosons in association with a hard jet.
           The next-to-leading-order corrections are included for 
           vector boson pair production~(VV)
           and the production of a vector boson in association with the Higgs boson~(V+Higgs)
           as well as in the decay of the Higgs boson, $h^0 \rightarrow b \,\bar{b}$.
           A SM Higgs boson was assumed with a mass of 115 $\mathrm{GeV}$.}
  \label{fig:nloPlots}
\end{figure}
\begin{table}[t]
  \begin{center}
{
\renewcommand{\arraystretch}{1.2}
    \begin{tabular}{|c| c| c| c|}
      \hline 
      \multicolumn{4}{|c|}{Significance} \\ 
      \hline \hline
      \multirow{1}{*}{Process} & Order  & $\tfrac{S}{\sqrt{B}}$ & $\tfrac{S}{\sqrt{B}}$\\
        \multirow{1}{*}{} & & \textsf{Herwig++} default & \textsf{Herwig++} tune\\
      \hline
      \multirow{2}{*}{$Z^0 \rightarrow l^+l^-$} & LO  & 1.17 & $1.24^{+0.36}_{-0.11}$\\
      & NLO & 1.57 & $1.96^{+0.29}_{-0.30}$\\ \hline
      \multirow{2}{*}{$Z^0 \rightarrow \nu\,\bar{\nu}$} & LO  & 2.18 & $2.89^{+0.19}_{-0.60}$\\
      & NLO & 2.95 & $4.04^{+0.25}_{-0.90}$ \\ \hline
      \multirow{2}{*}{$W \rightarrow l\,\nu$} & LO  & 1.88 & $2.32^{+0.15}_{-0.27}$\\
      & NLO & 2.63 & $3.20^{+0.29}_{-0.36}$\\ \hline
      \multirow{2}{*}{Total} & LO  & 2.98 & $3.71^{+0.29}_{-0.53}$\\
      & NLO & 4.09 & $5.20^{+0.43}_{-0.81}$ \\ \hline
    \end{tabular}
}
    \caption{The significance of the different processes for the 
    leading- and next-to-leading-order matrix elements. The significance is
    calculated using all masses in the range 112-120 $\mathrm{GeV}$.}
    \label{tab:StoB}
  \end{center} 
\end{table}

In addition the presence of a hard jet with $p_{T_j} > p_T^{\min}$ with
substructure is required.
The substructure analysis of Ref.~\cite{Butterworth:2008iy} proceeds with the 
hard jet $j$ with some radius $R_j$, a mass $m_j$ and in a mass-drop algorithm:
\begin{enumerate}
  \item the two subjets which were merged to form the jet, ordered such that
  the mass of the first jet $m_{j_1}$ is greater than that of the second
  jet $m_{j_2}$, are obtained;
    \label{MD:one}
  \item if $m_{j_1} < \mu \,m_j$ and
\begin{equation}
y = \frac{\min(p_{Tj_1}^2,\,p_{Tj_2}^2)}{m_j^2}\Delta R^2_{j_1,j_2} > y_{\mathrm{cut}},
\end{equation}
where $\Delta R^2_{j_1,j_2}=(y_{j_1}-y_{j_2})^2+(\phi_{j_1}-\phi_{j_2})^2$, and 
$p_{Tj_{1,2}}$, $\eta_{j_{1,2}}$, $\phi_{j_{1,2}}$ are
the transverse momenta, rapidities and azimuthal angles of jets 1 and 2,
respectively, then $j$ is in the heavy particle region.
If the jet is not in the heavy particle region the procedure is repeated using
the first jet.
\end{enumerate}
This algorithm requires that $j_{1,2}$ are $b$-tagged and takes $\mu =
0.67$ and \mbox{$y_{\mathrm{cut}} = 0.09$}.
A uniform $b$-tagging efficiency of 60\% was used
with a uniform mistagging probability of 2\%.
 The heavy jet selected by this
procedure is considered to be the Higgs boson candidate jet. Finally,
there is a filtering procedure on the Higgs boson candidate jet,
$j$. The jet, $j$, is resolved on a finer scale by setting a new
radius $R_{\mathrm{filt}}=\mathrm{min}(0.3,R_{b\bar{b}}/2)$, where
from the previous mass-drop condition, $R_{b\bar{b}} = \Delta
R^2_{j_1,j_2}$. The three hardest subjects of this filtering process
are taken to be the Higgs boson decay products, where the two hardest
are required to be $b$-tagged.

All three analyses require that:
\begin{itemize}
\item after the reconstruction of the vector boson, there are 
  no additional leptons with pseudorapidity $|\eta| < 2.5$ and $p_T > 30 \, \mathrm{GeV}$;
  \item other than the Higgs boson candidate, there are
  no additional $b$-tagged jets with pseudorapidity $|\eta| < 2.5$ and $p_T > 50 \, \mathrm{GeV}$.
\end{itemize} 
In addition, due to
top contamination, criterion (\ref{analysis:W}) requires that other
than the Higgs boson candidate, there are no additional $b$-tagged jets with
$|\eta| < 3$ and \mbox{$p_T > 30 \, \mathrm{GeV}$}. For all events, the candidate
Higgs boson jet should have $p_T > p_T^{\mathrm{min}}$. The analyses were implemented
using the \textsf{Rivet} system \cite{Buckley:2010ar}.

\begin{figure}[t!!]
  \begin{center}
    \subfigure[Selection criterion (a)]{
      \includegraphics[width=0.45\textwidth]{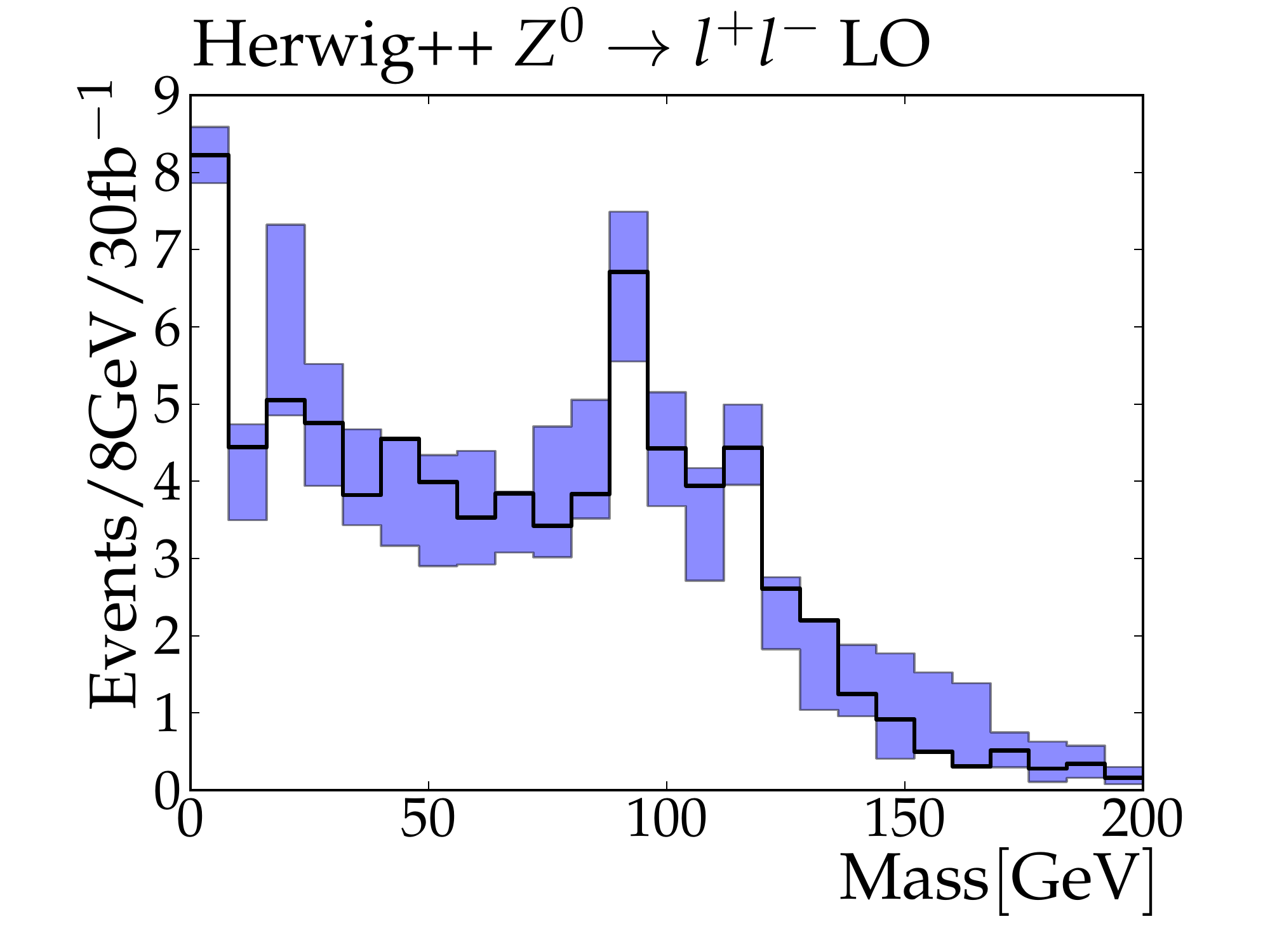}
    }  
    \subfigure[Selection criterion (b)]{
      \includegraphics[width=0.45\textwidth]{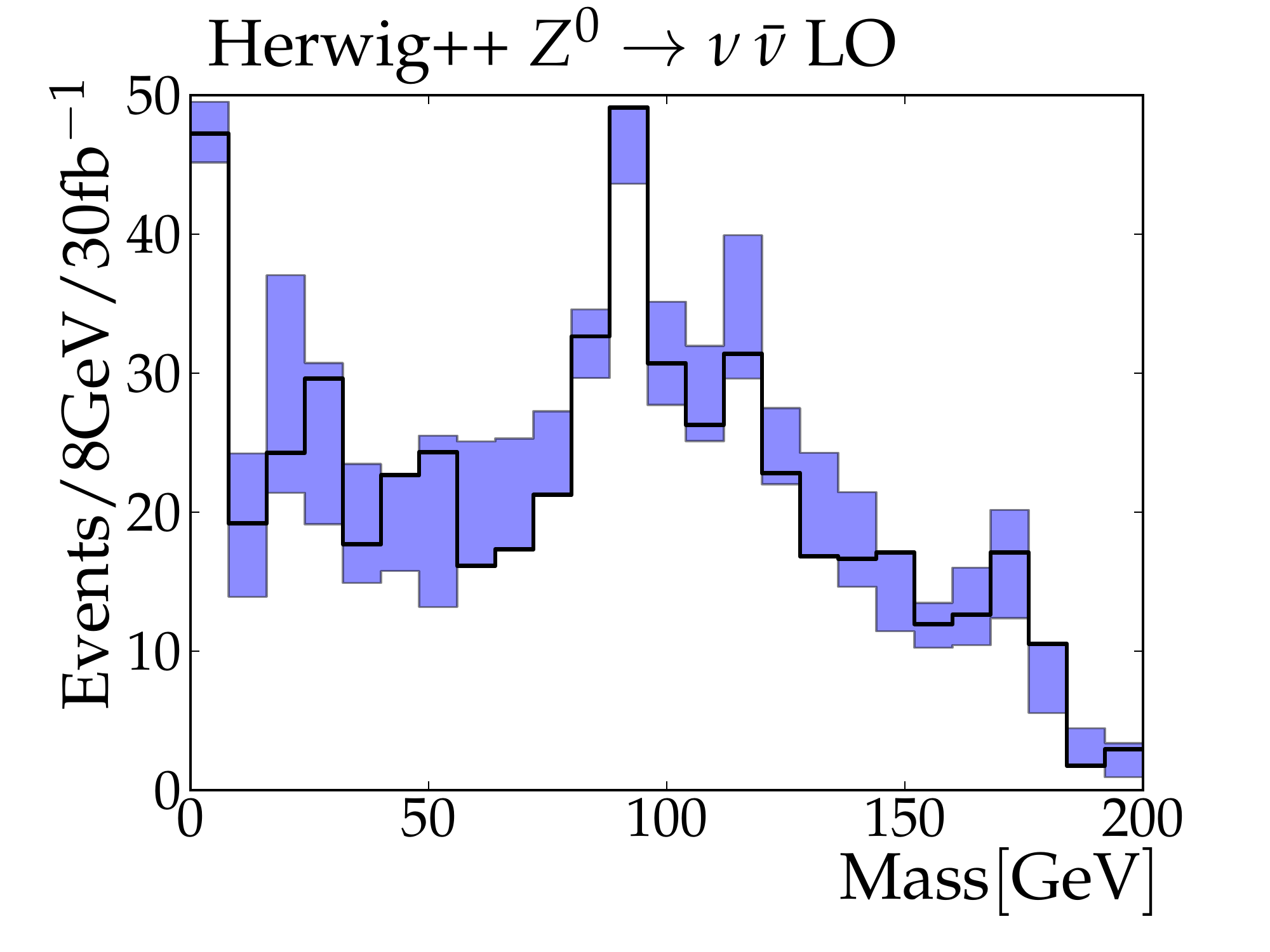}
    }
    \subfigure[Selection criterion (c)]{
      \includegraphics[width=0.45\textwidth]{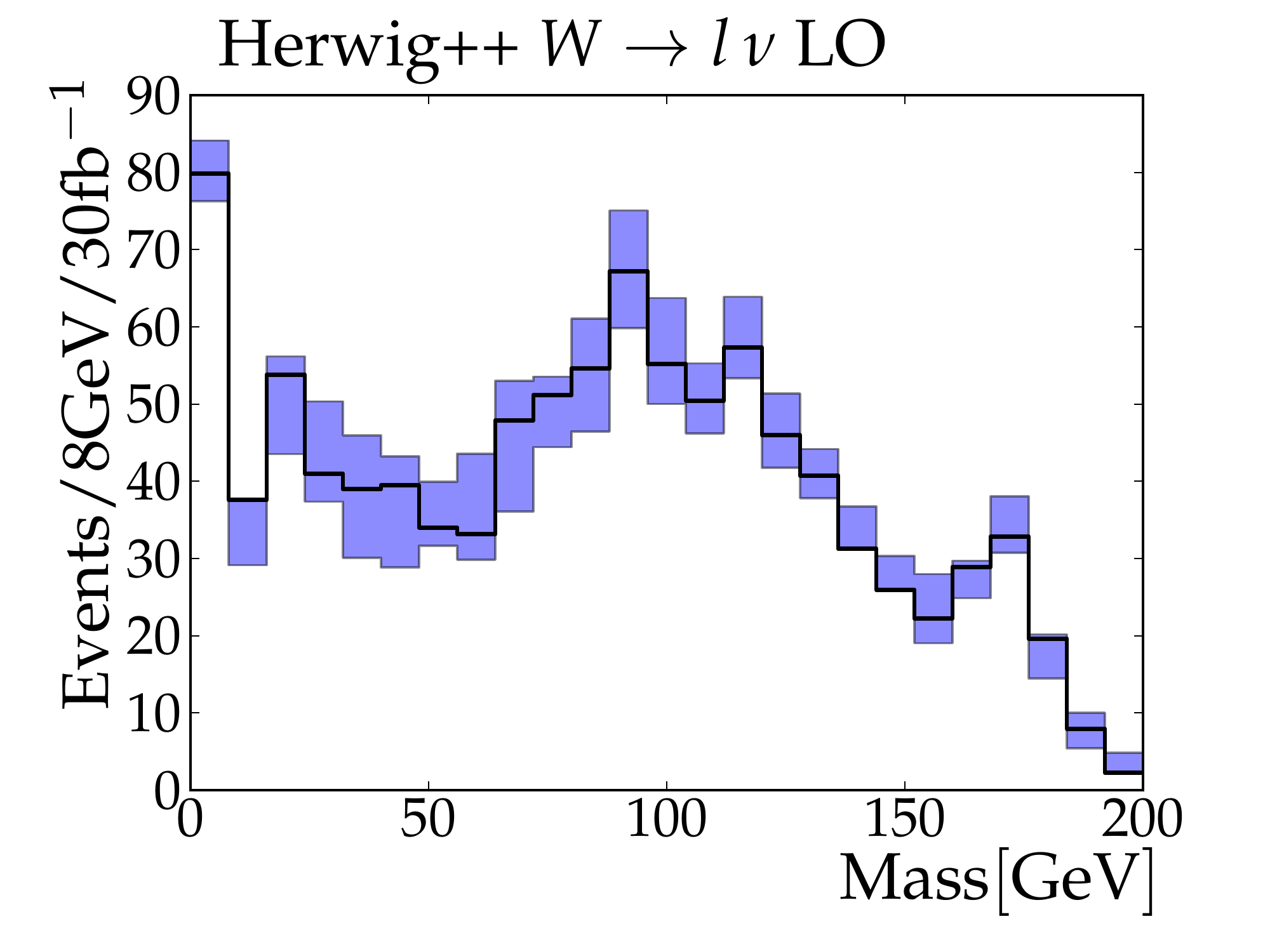}
    }
    \subfigure[Sum of criteria (a), (b) and (c)]{
      \includegraphics[width=0.45\textwidth]{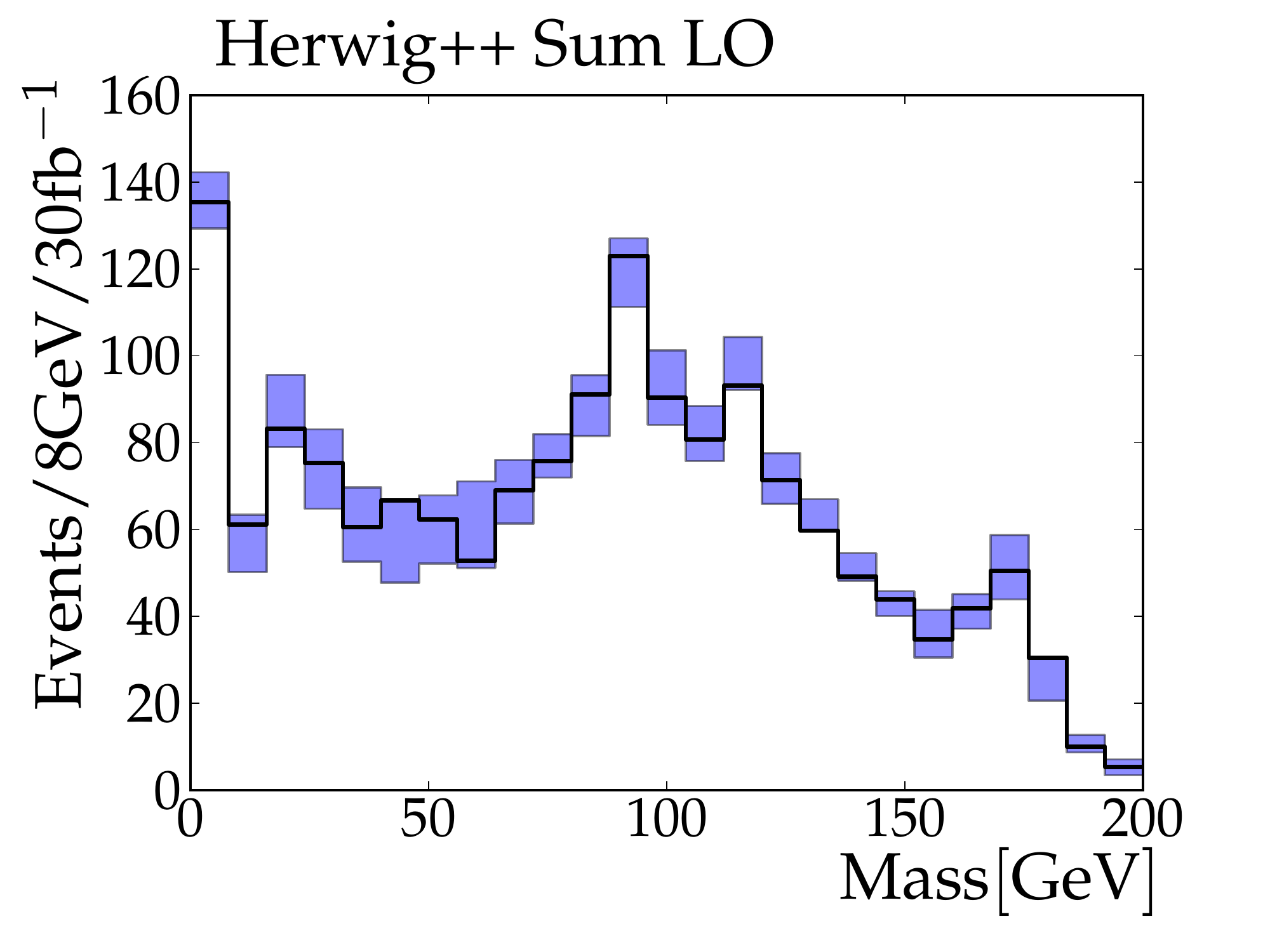}
    }
  \end{center}
  \caption{Results for the reconstructed Higgs boson mass distribution using leading-order matrix elements.
           A SM Higgs boson was assumed with a mass of 115 $\mathrm{GeV}$. The envelope shows
           the uncertainty from the Monte Carlo simulation.}
  \label{fig:loPlotsEnvelope}
\end{figure}
\begin{figure}[t!!]
  \begin{center}
    \subfigure[Selection criterion (a)]{
      \includegraphics[width=0.45\textwidth]{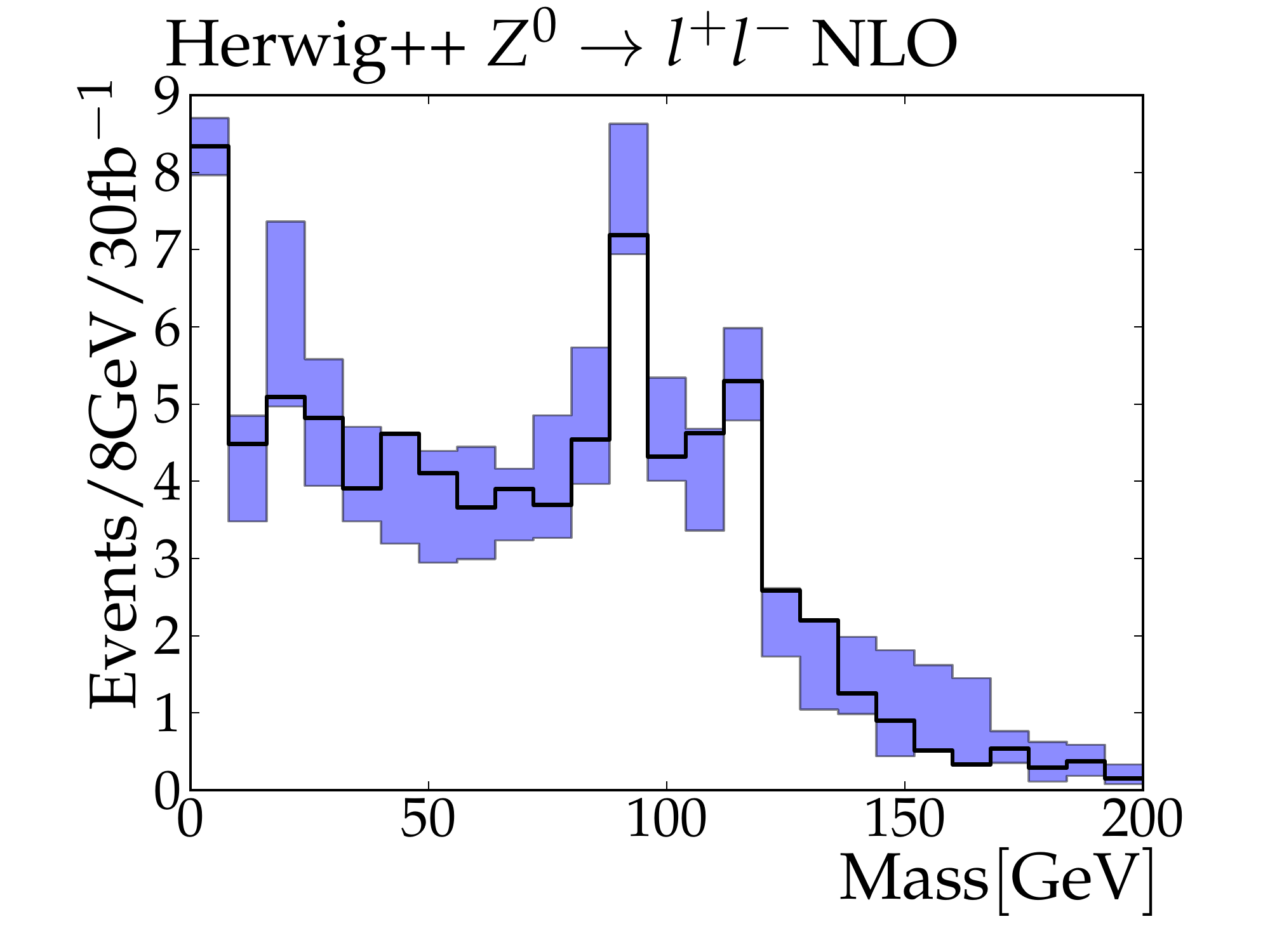}
    }  
    \subfigure[Selection criterion (b)]{
      \includegraphics[width=0.45\textwidth]{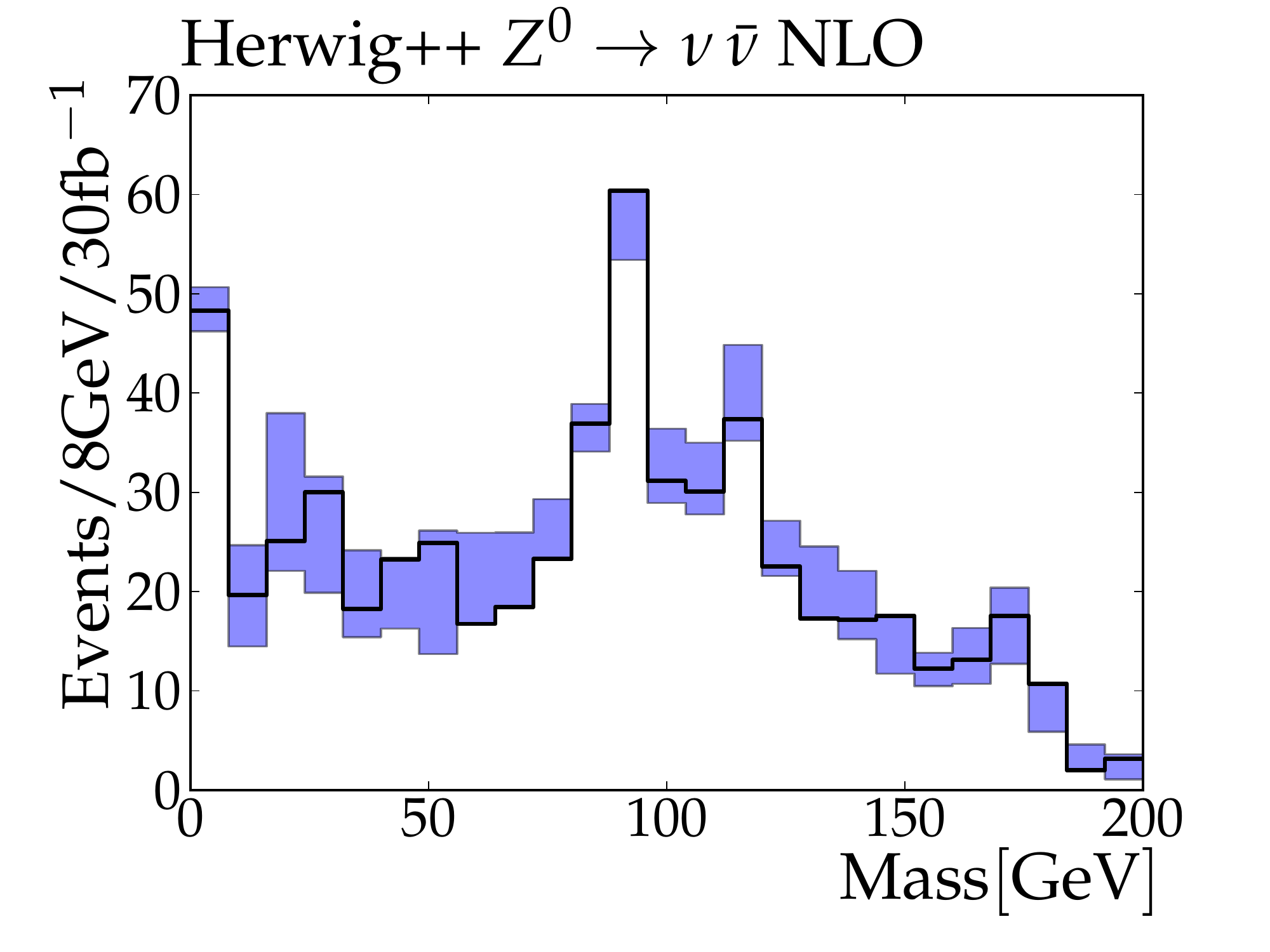}
    }
    \subfigure[Selection criterion (c)]{
      \includegraphics[width=0.45\textwidth]{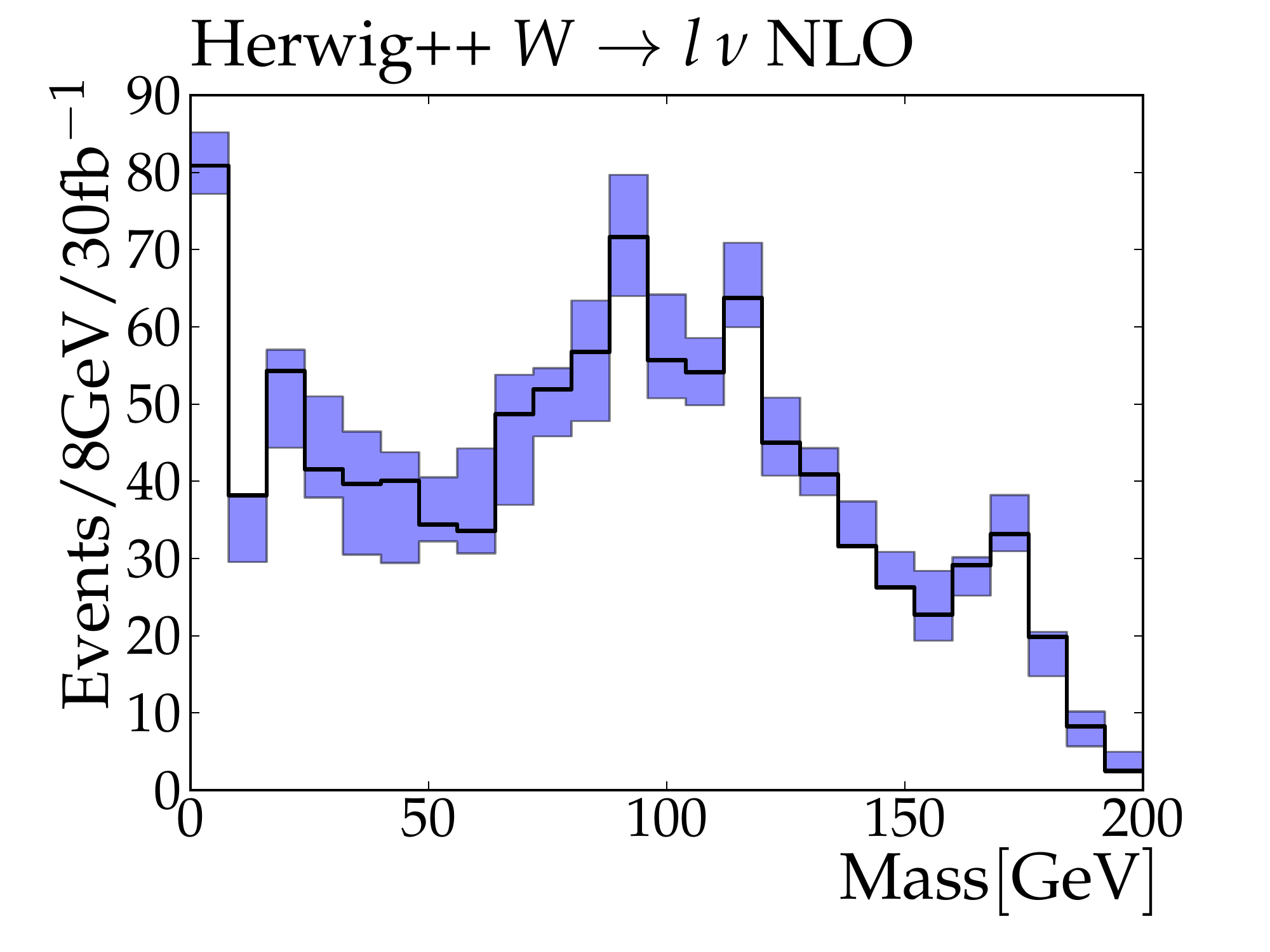}
    } 
    \subfigure[Sum of criteria (a), (b) and (c)]{
      \includegraphics[width=0.45\textwidth]{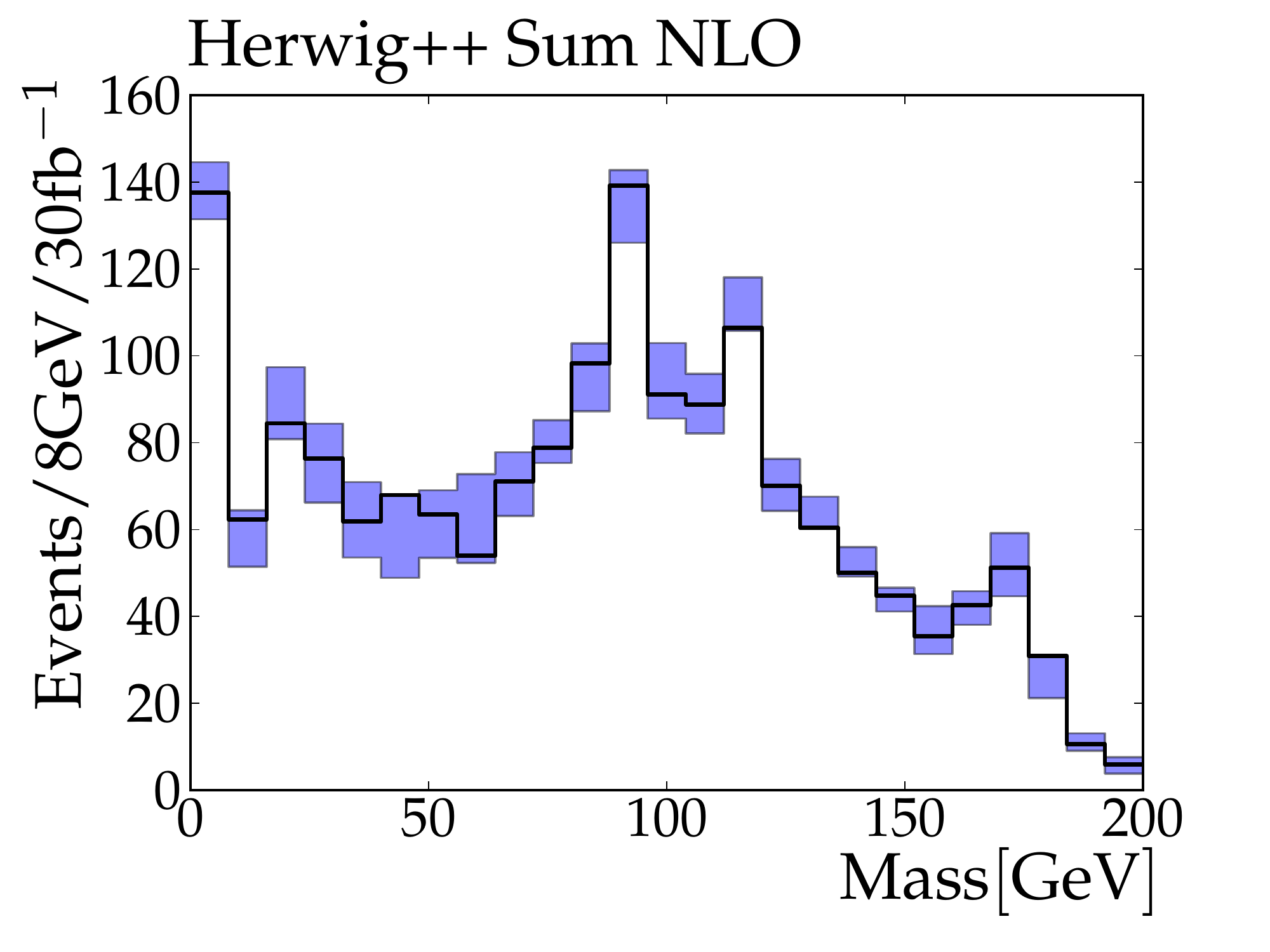}
    }
  \end{center}
  \caption{Results for the reconstructed Higgs boson mass distribution using
           leading-order matrix elements for  top quark pair production, and
           the production of $W^\pm$ and $Z^0$ bosons in association with a hard jet.
           The next-to-leading-order corrections are included for 
           vector boson pair production
           and the production of a vector boson in association with the Higgs boson
           as well as in the decay of the Higgs boson, $h^0 \rightarrow b \,\bar{b}$.
           A SM Higgs boson was assumed with a mass of 115 $\mathrm{GeV}$.
           The envelope shows the uncertainty from the Monte Carlo simulation.}
  \label{fig:nloPlotsEnvelope}
\end{figure}
The plots shown in Fig.~\ref{fig:loPlots} use the leading-order matrix
elements for the production and decay of Higgs boson but
the $W$, $Z$ and top \cite{Hamilton:2006ms} have matrix element
corrections for their decays. The plots shown in
Fig.~\ref{fig:nloPlots} have leading-order $t\bar{t}$ production,
leading-order vector boson plus jet production (with the same matrix element
corrections as the LO matrix elements) but the NLO vector boson pair
production~\cite{Hamilton:2010mb} and NLO vector and Higgs boson associated
production~\cite{Hamilton:2009za}.
In addition we have implemented the corrections to the decay $h^0 \rightarrow b \,\bar{b}$
in the POWHEG scheme, as described in Appendix~\ref{app:powheg}.
The signal significances are outlined in Table~\ref{tab:StoB}.

The uncertainties due to the Monte Carlo simulation are shown as bands in 
Figs.\,\ref{fig:loPlotsEnvelope}~and~\ref{fig:nloPlotsEnvelope}. 
As there are correlations between the different processes the uncertainty is determined
for the sum of all processes. Whilst it would be possible to show the
envelope for the individual processes, this would not offer any
information on the envelope for the sum of the processes which is the
result of interest. In addition the uncertainty on the significance is
shown in Table~\ref{tab:StoB}.

\section{Conclusions}

Monte Carlo simulations are an essential tool in the analysis of modern
collider experiments. While significant effort has been devoted to 
the tuning of the parameters to produce a best fit there has been much less 
effort understanding the uncertainties in these results. In this paper
we have produced a set of tunes which can be used to assess this uncertainty
using the \textsf{Herwig++} Monte Carlo event generator.

We then used these tunes to assess the uncertainties on the mass-drop analysis 
of Ref.~\cite{Butterworth:2008iy} using \textsf{Herwig++}
with both leading- and next-to-leading-order matrix elements including a POWHEG simulation of
the decay $h^0\rightarrow b \bar{b}$.

We find that while the jet substructure technique has significant 
potential as a Higgs boson discovery channel, we need to be confident of our tunes to
investigate this with Monte Carlo simulations.

The error tunes and procedure here can now be used in other
analyses where the uncertainty due to the Monte Carlo simulation is
important.

\section{Acknowledgements}

We are grateful to all the other members of the \textsf{Herwig++}
collaboration and Hendrik Hoeth for valuable discussions and the authors
of Ref.~\cite{Butterworth:2008iy} for help in reproducing their results.
We acknowledge the use of the
UK Grid for Particle Physics in producing the results. This work was
supported by the Science and Technology Facilities Council.
DW acknowledges support by the STFC studentship ST/F007299/1.

\appendix
\section{Observables and weights used to tune \textsf{Herwig++}}
\label{app:observables}
The weights and observables used in the \textsf{Professor} tuning
system are outlined in the tables below.
\begin{table}[h!!!]
  \footnotesize
  \begin{center}
    \begin{tabular}{|l| c|l| c|}
      \hline
      Observable & Weight & Observable & Weight \\
      \hline \hline
       $K^{*\pm}(892)$ spectrum & 1.0 &
       $\Lambda^0$ spectrum & 1.0 \\
       $\rho$ spectrum & 1.0 &
       $\pi^0$ spectrum & 1.0 \\
       $\omega(782)$ spectrum & 1.0 &
       $p$ spectrum & 1.0 \\
       $\Xi^-$ spectrum & 1.0 &
       $\eta'$ spectrum & 1.0 \\
       $K^{*0}(892)$ spectrum & 1.0 &
       $\Xi^0(1530)$ spectrum & 1.0 \\
       $\phi$ spectrum & 1.0 &
       $\pi^\pm$ spectrum & 1.0 \\
       $\Sigma^\pm(1385)$ spectrum & 1.0 &
       $\eta$ spectrum & 1.0 \\
       $\gamma$ spectrum & 1.0 &
       $K^0$ spectrum & 1.0 \\
       $K^\pm$ spectrum & 1.0 && \\
    \hline
    \end{tabular}
    \caption{Observables used in the tuning and associated weights for observables taken 
             from~\cite{Barate:1996fi}.}
\label{tab:ALEPH_1996_S3486095}
\vspace{-3cm}
  \end{center}
\end{table}

\begin{table}[p!!!]
  \footnotesize
  \begin{center}
    \begin{tabular}{|l| c|}
      \hline
      Observable & Weight \\
      \hline \hline
       Sphericity, $S$ & 1.0 \\
       Energy-energy correlation, EEC & 1.0 \\
       Aplanarity, $A$ & 2.0 \\
       Mean out-of-plane $p_\perp$ in GeV w.r.t. thrust axes vs. $x_p$ & 1.0 \\
       Mean charged multiplicity & 150.0 \\
       Mean $p_\perp$ in GeV vs. $x_p$ & 1.0 \\
       Planarity, $P$ & 1.0 \\
       Thrust major, $M$ & 1.0 \\
       Oblateness = $M - m$ & 1.0 \\
       Out-of-plane $p_\perp$ in GeV w.r.t. sphericity axes & 1.0 \\
       $D$ parameter & 1.0 \\
       $1-\text{Thrust}$ & 1.0 \\
       Out-of-plane $p_\perp$ in GeV w.r.t. thrust axes & 1.0 \\
       Log of scaled momentum, $\log(1/x_p)$ & 1.0 \\
       In-plane $p_\perp$ in GeV w.r.t. sphericity axes & 1.0 \\
       In-plane $p_\perp$ in GeV w.r.t. thrust axes & 1.0 \\
       Thrust minor, $m$ & 2.0 \\
       $C$ parameter & 1.0 \\
       Scaled momentum, $x_p = |p|/|p_\text{beam}|$ & 1.0 \\
    \hline
    \end{tabular}
    \caption{Observables used in the tuning and associated weights for observables taken 
             from~\cite{Abreu:1996na}.}
\label{tab:DELPHI_1996_S3430090}
  \end{center}
  \footnotesize
  \begin{center}
    \begin{tabular}{|l| c|l| c|}
      \hline
      Observable & Weight & Observable & Weight \\
      \hline \hline
       Mean $\rho^0(770)$ multiplicity & 10.0 &
       Mean $\chi_{c1}(3510)$ multiplicity & 10.0 \\
       Mean $\Delta^{++}(1232)$ multiplicity & 10.0 &
       Mean $D^+$ multiplicity & 10.0 \\
       Mean $K^{*+}(892)$ multiplicity & 10.0 &
       Mean $\Sigma^+$ multiplicity & 10.0 \\
       Mean $\Sigma^0$ multiplicity & 10.0 &
       Mean $f_1(1285)$ multiplicity & 10.0 \\
       Mean $\Lambda_b^0$ multiplicity & 10.0 &
       Mean $f_2(1270)$ multiplicity & 10.0 \\
       Mean $K^+$ multiplicity & 10.0 &
       Mean $J/\psi(1S)$ multiplicity & 10.0 \\
       Mean $\Xi^0(1530)$ multiplicity & 10.0 &
       Mean $B^+_u$ multiplicity & 10.0 \\
       Mean $\Lambda(1520)$ multiplicity & 10.0 &
       Mean $B^**$ multiplicity & 10.0 \\
       Mean $D^{*+}_s(2112)$ multiplicity & 10.0 &
       Mean $\Lambda_c^+$ multiplicity & 10.0 \\
       Mean $\Sigma^-(1385)$ multiplicity & 10.0 &
       Mean $D^0$ multiplicity & 10.0 \\
       Mean $f_1(1420)$ multiplicity & 10.0 &
       Mean $f_2'(1525)$ multiplicity & 10.0 \\
       Mean $\phi(1020)$ multiplicity & 10.0 &
       Mean $\Sigma^\pm$ multiplicity & 10.0 \\
       Mean $K_2^{*0}(1430)$ multiplicity & 10.0 &
       Mean $D_{s2}^+$ multiplicity & 10.0 \\
       Mean $\Omega^-$ multiplicity & 10.0 &
       Mean $K^{*0}(892)$ multiplicity & 10.0 \\
       Mean $\Sigma^\pm(1385)$ multiplicity & 10.0 &
       Mean $\Sigma^-$ multiplicity & 10.0 \\
       Mean $\psi(2S)$ multiplicity & 10.0 &
       Mean $\pi^+$ multiplicity & 10.0 \\
       Mean $D^{*+}(2010)$ multiplicity & 10.0 &
       Mean $f_0(980)$ multiplicity & 10.0 \\
       Mean $B^*$ multiplicity & 10.0 &
       Mean $\Sigma^+(1385)$ multiplicity & 10.0 \\
       Mean $\pi^0$ multiplicity & 10.0 &
       Mean $D^+s$ multiplicity & 10.0 \\
       Mean $\eta$ multiplicity & 10.0 &
       Mean $p$ multiplicity & 10.0 \\
       Mean $a_0^+(980)$ multiplicity & 10.0 &
       Mean $B^0_s$ multiplicity & 10.0 \\
       Mean $D_{s1}^+$ multiplicity & 10.0 &
       Mean $K^0$ multiplicity & 10.0 \\
       Mean $\rho^+(770)$ multiplicity & 10.0 &
       Mean $B^+, B^0_d$ multiplicity & 10.0 \\
       Mean $\Xi^-$ multiplicity & 10.0 &
       Mean $\Lambda$ multiplicity & 10.0 \\
       Mean $\omega(782)$ multiplicity & 10.0 &
       Mean $\eta'(958)$ multiplicity & 10.0 \\
       Mean $\Upsilon(1S)$ multiplicity & 10.0 && \\
    \hline
    \end{tabular}
    \caption{Multiplicities used in the tuning and associated weights for observables taken 
             from~\cite{Amsler:2008zzb}.}
\label{tab:PDG_HADRON_MULTIPLICITIES}
  \end{center}
\end{table}

\begin{table}[t]
  \footnotesize
  \begin{center}
    \begin{tabular}{|l| c|}
      \hline
      Observable & Weight \\
      \hline \hline
       $b$ quark fragmentation function $f(x_B^\text{weak})$ & 7.0 \\
       Mean of $b$ quark fragmentation function $f(x_B^\text{weak})$ & 3.0 \\
    \hline
    \end{tabular}
    \caption{Observables used in the tuning and associated weights for observables taken 
             from~\cite{Barker:994376}.}
  \label{tab:DELPHI_2002_069_CONF_603}
  \end{center}
\end{table}
\section{Simulation of \boldmath{$h^0 \rightarrow b \bar{b}$} using the POWHEG Method}
\label{app:powheg}

The NLO  differential decay rate in the POWHEG~\cite{Nason:2004rx} approach is
\begin{equation}
  \label{eq:POWHEGdiffXS}
  \mathrm{d}\sigma =
  \bar{B}(\Phi_m)\mathrm{d}\Phi_\mathrm{B}\Biggl[\Delta_\mathrm{R}^\mathrm{NLO}(0)
    +  \Delta_\mathrm{R}^\mathrm{NLO}(p_T)\frac{R(\Phi_m,\Phi_\mathrm{ 1})}{B(\Phi_m)}\mathrm{d}\Phi_\mathrm{1}
    \Biggr],
\end{equation}
where
\begin{equation}
  \label{eq:BbarPOWHEG}
  \bar{B}(\Phi_m) = B(\Phi_m) + V(\Phi_m) + \int \left( R(\Phi_m,
  \Phi_\mathrm{ 1}) - \sum\limits_i D_i(\Phi_m, \Phi_\mathrm{1}) \right)
  \mathrm{d}\Phi_\mathrm{1}\mathrm{.}
\end{equation}
Here $B(\Phi_m)$ is the
leading-order Born differential decay rate, $V(\Phi_m)$ the
regularized virtual contribution, $D_i(\Phi_m, \Phi_\mathrm{1})$ the
counter terms regularizing the real emission and
$R(\Phi_m,\Phi_\mathrm{1})$ the real emission contribution.
The leading-order process has $m$ outgoing partons, with associated phase space
$\Phi_m$. The virtual and
Born contributions depend only on this $m$-body phase space.
The real emission phase space, $\Phi_{m+1}$, is factorised
into the $m$-body phase space and the phase space, $\Phi_\mathrm{1}$, 
describing the radiation of an extra parton.

The Sudakov form factor in the POWHEG method is
\begin{equation}
  \label{eq:powhegSudakov}
  \Delta_\mathrm{R}^\mathrm{NLO} = \exp \Biggl[ - \int \mathrm{d}\Phi_\mathrm{1} \frac{R(\Phi_m, \Phi_\mathrm{1})}{B(\Phi_m)} 
    \theta (k_T(\Phi_m,\Phi_\mathrm{1}) -p_T) \Biggr],
\end{equation}
where $k_T(\Phi_m,\Phi_\mathrm{1})$ is the transverse momentum of the emitted parton. 

In order to implement the decay of the Higgs boson in the POWHEG scheme in
\textsf{Herwig++} we need to generate the Born configuration according to 
Eq.\,\ref{eq:BbarPOWHEG} and the subsequent hardest emission according
to Eq.\,\ref{eq:powhegSudakov}. The generation of the truncated and vetoed 
parton showers from these configurations then proceeds as described in Refs.\cite{Bahr:2008pv,Hamilton:2008pd,Hamilton:2009za,D'Errico:2011um}.

\begin{figure}[t!!]
  \begin{center}
      \includegraphics[width=0.45\textwidth]{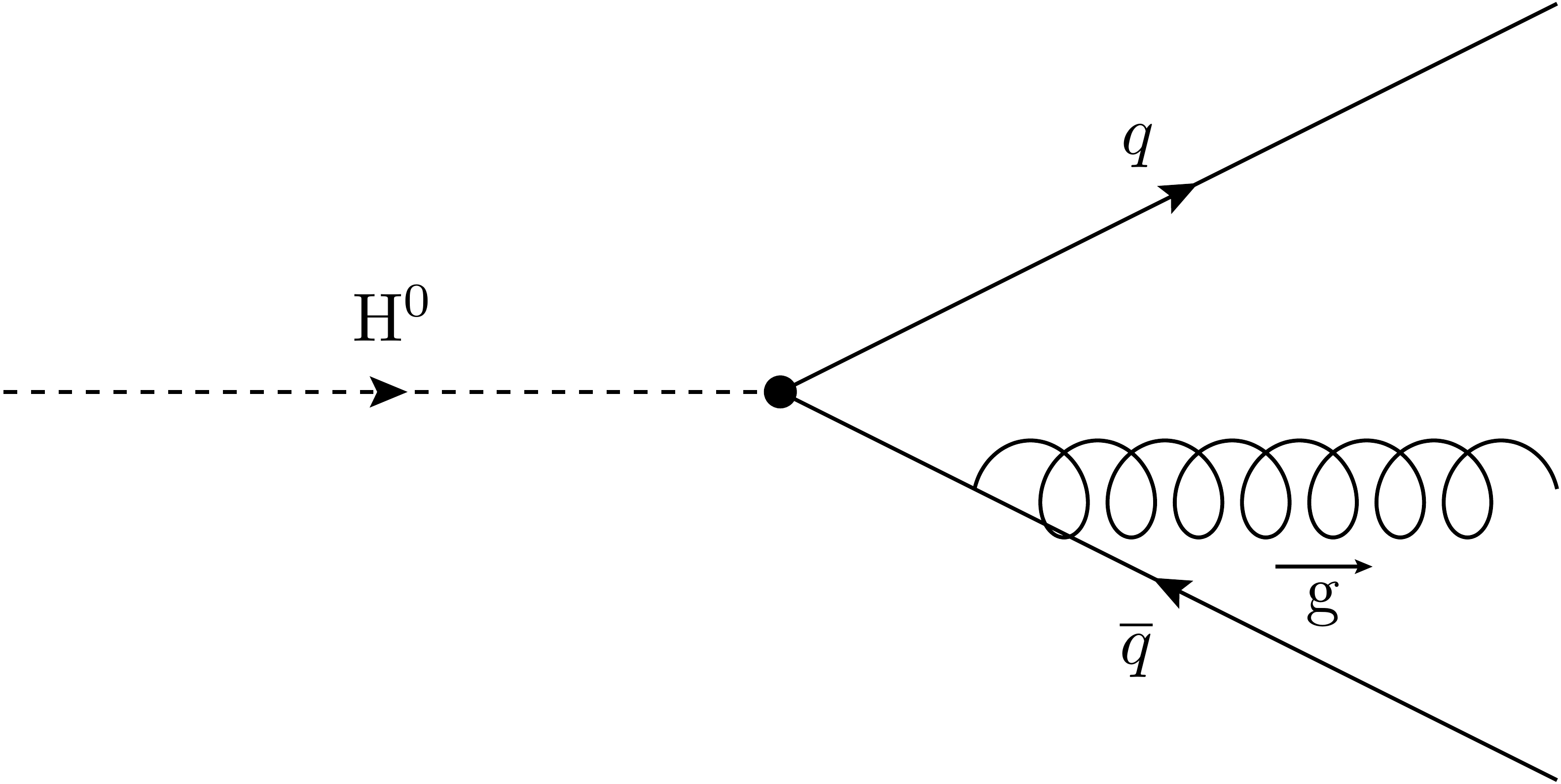}\hfill
      \includegraphics[width=0.45\textwidth]{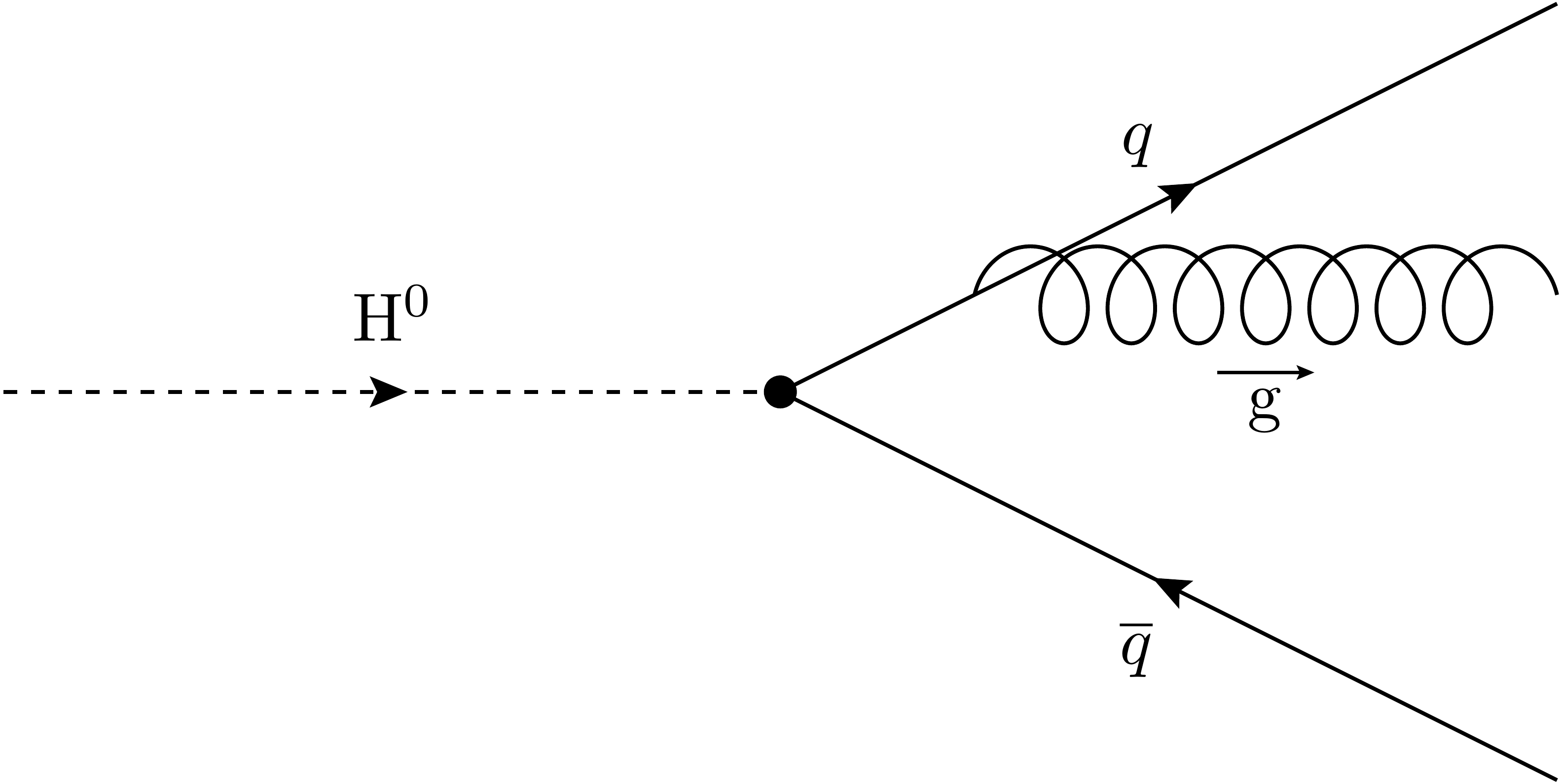}
  \end{center}
  \caption{The two real-emission processes contributing to the NLO decay rate.}
  \label{fig:realEmission}
\end{figure}
The virtual contribution for {$h^0 \rightarrow b \bar{b}$ was calculated in
Ref.~\cite{Braaten:1980yq}. The corresponding real
emission contribution, see Fig.~\ref{fig:realEmission}, is
\label{eq:realEmission}
\begin{align}
  |\mathcal{M}_R|^2 =& \,|\mathcal{M}_2|^2\frac{\mathrm{C_F}8\pi\alpha_\mathrm{s}}{M^2_H(1-4\mu^2)} 
  \Biggl[ 2 + \frac{1-x_q}{1-x_{\bar{q}}}  + \frac{(8\mu^4-6\mu^2+1)}{(1-x_q)(1-x_{\bar{q}})}\nonumber\\
    &\, -2(1-4\mu^2)\frac{1}{1-x_q}
    -2\mu^2(1-4\mu^2)\frac{1}{(1-x_q)^2} + (x_q \leftrightarrow
  x_{\bar{q}})\Biggr],
\end{align}
where $\mathcal{M}_2$ is the leading-order matrix element,
$\mathrm{C_F} = \tfrac{4}{3}$, $m_q$ is the mass of the bottom quark, $M_H$ is 
the mass of the Higgs boson, $\mu = \tfrac{m_q}{M_H}$ and 
$x_i =\tfrac{2E_i}{M_H}$. We use the Catani-Seymour
subtraction scheme \cite{Catani:2002hc} where the counter terms are
\begin{align}
  D_{i} &= \mathrm{C_F}\frac{8\pi\alpha_S}{s}|\mathcal{M}_2|^2 \nonumber\\
  &\times
  \frac1{1-x_j}
  \left\{
  \frac{2(1-2\mu^2)}{2-x_i-x_j}
  -\sqrt{\frac{1-4\mu^2}{x_j^2-4\mu^2}}
  \frac{x_j-2\mu^2}{1-2\mu^2}
  \left[2 + \frac{x_i - 1}{x_j-2\mu^2}
    +\frac{2\mu^2}{1-x_j}
    \right]\right\},
\end{align}
where for $D_i$, $i$ is the emitting parton and $j$ is the
spectator parton. In practice, as the counter terms can become
negative in some regions, we use
\begin{equation}
  R(\Phi_m,\Phi_\mathrm{1}) - \sum\limits_i D_i(\Phi_m, \Phi_\mathrm{1})
  = \sum\limits_i \Biggl[ \frac{R(\Phi_m,\Phi_\mathrm{1}) \left|D_i(\Phi_m, \Phi_\mathrm{1})\right|}{\sum\limits_j\left|D_j(\Phi_m, \Phi_\mathrm{1})\right|}
    - D_i(\Phi_m, \Phi_\mathrm{1}) \Biggr].
\end{equation}
We have also regulated
singularities in the virtual term $V(\Phi_m)$ with the integrated counter
terms from the Catani-Seymour subtraction scheme allowing us to generate
the Born configuration according to $\bar{B}(\Phi_m)$.

The hardest emission for each leg is generated according to
\begin{align}
  \label{eq:pTSudakovFormFactor}
  \nonumber
  \Delta_{i\,\mathrm{R}}^\mathrm{NLO} = \exp \Biggl[& 
    - \frac{M_H^2}{16\pi^2(1-4\mu^2)^{\frac{1}{2}}} \times \\
    &\int \mathrm{d}\,x_1 \,\mathrm{d}\,x_2\, \mathrm{d}\,\phi \,
    \frac{R(\Phi_{m+1})}{B(\Phi_m)}\frac{|D_i|}{\sum_j|D_j|}
    \theta (k_T(\Phi_m,\Phi_\mathrm{1}) -p_T) \Biggr].
\end{align}
However this form is not suitable for the generation of the hardest emission.
Instead we perform a Jacobian transformation and use the transverse momentum, $p_T$,
rapidity, $y$, and azimuthal angle, $\phi$, of the radiated gluon to define the
phase space $\Phi_1$.

The momenta of the Higgs boson decay products are
\begin{subequations}
\begin{align}
  \label{eq:p1}
  p_1 &= \frac{M_H}{2}\biggl(x_1;-x_{\perp}\cos(\phi),-x_{\perp}\sin(\phi),\pm\sqrt{x_1^2-x_{\perp}^2-4\mu^2}\biggr),\\
  \label{eq:p2}
  p_2 &= \frac{M_H}{2}\biggl(x_2;0,0,-\sqrt{x_2^2-4\mu^2}\biggr),\\
  \label{eq:p3}
  p_3 &= \frac{M_H}{2}\biggl(x_3;x_{\perp}\cos(\phi),x_{\perp}\sin(\phi),\pm\sqrt{x_3^2-x_{\perp}^2}\biggr),
\end{align}
\end{subequations}
where partons~${1,2,3}$ are the
radiating bottom quark, spectator antibottom quark and radiated gluon, 
respectively. The energy fractions $x_i=\tfrac{2E_i}{M_H}$ and\linebreak \mbox{$x_\perp=\tfrac{2p_T}{M_H}$}.
Using the conservation of momentum in the $z$-direction
and \linebreak \mbox{$x_1+x_2+x_3=2$} gives
\begin{align}
x_\perp^2=\left( 2-x_{{1}}-x_2 \right) ^2-{\frac { \left( -2+2x_{{1}}+2
x_2-x_2x_{{1}}-{x_2}^2 \right) ^2}{{x_2}^2-4\mu^2}}.
\end{align}
Together with the definition, $x_3 = x_{\perp} \cosh y$,
we obtain the Jacobian
\begin{equation}
\label{eq:Jacobian}
\biggl|\frac{\partial x_1\partial x_2}{\partial p_T\partial y}\biggr| = 
\frac{x_{\perp}}{M_H}\frac{x_{\perp}(x_2^2-4\mu^2)^{\frac{3}{2}}}{(x_1x_2-2\mu^2(x_1+x_2)+x_2^2-x_2)},
\end{equation}
for the transformation of the radiation variables.

 We can then generate the additional radiation according to 
Eq.\,\ref{eq:pTSudakovFormFactor} using the veto algorithm~\cite{Sjostrand:2006za}.
To achieve this we use an overestimate of the integrand in the Sudakov form factor,
$f(p_T) = \tfrac{c}{p_T}$, where $c$ is a suitable constant.
We first generate an emission according to
\begin{equation}
  \label{eq:overSudakov} 
  \Delta^{\mathrm{over}}_R(p_T) = \exp \Biggl[ 
    - \int_{p_T^{ }}^{p_T^{\max}} \int_{y_{\min}}^{y_{\max}} 
    \mathrm{d}\,p_T \,\mathrm{d}\,y\frac{c}{p_T} \Biggr],
\end{equation} 
using this overestimate, where $y_{\max} =\cosh^{-1}\left(\tfrac{M_H}{2p_T^{\min}}\right)$,
 $y_{\min} = -y_{\max}$, $p_T^{\max}$ is the maximum possible transverse 
momentum of the gluon and $p_T^{\min}$ is a parameter set in
the model, taken to be $1\,\mathrm{GeV}$.

The trial value of the transverse momentum is obtained by
solving \mbox{$\mathcal{R} = \Delta_R^{\mathrm{over}}$},
where $\mathcal{R}$ is a random number in $[0,1]$, {\it i.e.}
\begin{equation}
  \label{eq:trialEmissionpT}
  p_T= p_{T}^{\max}\,\mathcal{R}^{\frac{1}{c(y_{\max}-y_{\min})}}.
\end{equation}
Once the trial $p_T$ has been generated, $y$ and $\phi$ are also 
generated uniformly between $[y_{\min},y_{\max}]$ and $[0,2\pi]$,
respectively. The energy fractions of the
partons are obtained using the definition $x_3=x_\perp\cosh y$,
\begin{align}
  \label{eq:x1}
  x_1 = &\frac{1}{2(x_3-1)-\frac{x_{\perp}^2}{2}} 
  \biggl\{3x_3-2+\frac{x_{\perp}^2}{2}x_3-x_{\perp}^2-x_3^2 \nonumber\\
  &\pm\sqrt{(x_3^2 - x_{\perp}^2)((x_3-1)(4\mu^2+x_3-1)-\mu^2x_{\perp}^2)} \biggr\}.
\end{align}
and $x_2$ using energy conservation. As there
are two solutions for $x_1$ both solutions must be kept and used to
calculate the weight for a particular trial $p_T$. The signs of the 
$z$-components of the momenta are fixed by the sign of the rapidity
and momentum conservation.  Any momentum configurations
outside of the physically allowed phase space are rejected and a new
set of variables generated. The momentum configuration is accepted
with a probability given by the ratio of the true integrand to the
overestimated value. If the configuration is rejected, the procedure
continues with $p_T^{\max}$ set to the rejected $p_T$
until the trial value of $p_T$ is accepted or falls below the minimum
allowed value, $p_T^{\min}$. This procedure generates the
radiation variables correctly as shown in Ref.~\cite{Sjostrand:2006za}.

This procedure is used to generate a trial emission from both the bottom
and antibottom. The hardest potential emission is then selected
which correctly generates events according to Eq.\,\ref{eq:pTSudakovFormFactor}
using this competition algorithm.

\bibliography{Herwig++}
\end{document}